\documentclass{caosp308}

\usepackage{graphicx}
\usepackage{csquotes}

\usepackage{natbib}
\articleNo{301}
\pubyear{2021}
\volume{51}
\volnumber{1}
\firstpage{7}
\received{October 7, 2020}
\accepted{November 3, 2020}

\begin{document}

\hauthor{M.Yu.\,Skulskyy}

\title{Formation of magnetized spatial structures in the Beta Lyrae system }
\subtitle{III. Reflection of magnetically controlled matter in circumbinary structures in helium lines, in particular arising from metastable levels}

\author{M.Yu.\,Skulskyy
       }

\institute{
Lviv Polytechnic National University, Department of Physics, 79013, Lviv, Ukraine \email{mysky@polynet.lviv.ua}
          }

\date{October 7, 2020}

\maketitle

\begin{abstract}
Spatial gaseous structures in the Beta Lyrae system have been studied with the fact of change in the longitudinal component of the donor's magnetic field during the orbital period in mind. The investigation was based primarily on the study of the dynamics of the circumstellar structures surrounding the binary system as a whole. The special emphasis was placed on the study of complex helium lines, in particular those arising from metastable levels. A number of different observable facts from the ultraviolet to the red spectral region were analyzed. The configuration of the donor magnetic field is a factor that not only enhances mass transfer and influences the formation of spatial gas structures between stellar components but, to some extent, also affects the outflow of matter and the formation of external gas structures around this interacting binary system. Together with previous articles \citep{Skulskyy2020a, Skulskyy2020b}, the pieces of evidence of this work, confirming the reflection of magnetically controlled matter in circumbinary structures, define the basis for a coherent picture of the mass exchange between components and outflows of matter outwards.
\end{abstract}

\section{Introduction}
\label{intr}
In previous papers, the focus was on the research of the relationship between the structure of the donor magnetic field and its reflection in the characteristic physical parameters of the visible and infrared Beta Lyrae spectra \citep{Skulskyy2020a, Skulskyy2020b}. Taking into account the spatial configuration of the donor magnetic field, the emphasis was put on certain facts of its reflection in the moving magnetized accretions structures between the stellar components of this binary system. The data of absolute spectrophotometry and spectral analysis, changes along with the orbital phases of the radial velocities and intensities of complex emission-absorption lines were examined, primarily based on the own spectral and spectrophotometric observations in the visual range spectrum. This was due to a series of conscious and consistent spectral observations and relevant studies conducted mainly in 1980-1995, after the discovery of the donor magnetic field. Such a method of approach has already made it possible to make certain generalizations and reflect moving magnetized structures as the original phenomenon of mass transfer, which is inherent to this binary system.

It is clear that by the time when nothing was known about the donor's magnetic field and its possible effect on the physical conditions in the near and far gas structures of this binary system, a number of significant scientific studies had been completed. Some physical characteristics and certain parameters of radiating gas structures near both stellar components or in the common shell were often described by the known paradigms, which were later rejected or modified. This is especially true of researchers who have studied the spectral lines of the shell and the characteristics of the spectrum in the far ultraviolet region. A large number of scientific materials are available for their coverage and further rethinking. In this article, special attention is paid to the scientific works in order to investigate the influence of the magnetic field on the dynamics of the circumstellar structure surrounding this binary system as a whole. Clearly, to create a coherent picture, such studies should be conducted in comparison with the already obtained results of the investigation of the magnetized gas structures both near the donor and the gainer and between them.

The previous researches of \cite{Skulskyy2020a, Skulskyy2020b} demonstrate an efficient collision of magnetized plasma in the phases of observation of magnetic poles on the donor surface. These shock collisions with the accretion disk in the phases of the secondary quadrature, in which the high-temperature environment and the system of formed accretion flows are observed, are especially noted. There are obvious correlations between the phase variability of the donor magnetic field and the corresponding variability of dynamic and energy characteristics of different complex lines, but above all the strongest emission lines H$_{\alpha}$ and He\,I $\lambda$ 7065. Moreover, the phase boundaries of the location of the magnetic pole on the surface of the donor, above which the matter outflows are formed, were discovered. The additional loss of matter from the donor surface is observed mainly from the donor surface in the phase region round 0.855\,P, i.e., when the magnetic field pole is facing the gainer. This allowed us to predict certain correlations between the behavior of the above strong lines, as well as certain helium lines formed in shell gas structures extending outward from the binary system. Therefore, in this article, closer attention is paid to such lines of helium, primarily those arising from metastable levels.

It should be recalled that the donor's magnetic field changes significantly during the orbital period (which is close to 12.94\,d); phase changes in the range from zero to one are tied to the main eclipse of this binary system in the visible spectral region when the more massive accretor obscures of the bright donor; the schematic model of the Beta Lyrae system and the picture of mass transfer are shown in Fig.\ref{fig:1} of \cite{Skulskyy2020a}, which should be kept in mind for a better understanding of the geometry and physical properties of this interacting binary system in further analysis; as this article is a continuation of a number of ongoing comprehensive studies of the proposed topic, references to previous results are quite natural.

\section{Magnetic field and gaseous structures surrounding the binary system}
\subsection{Magnetic field and the He\,I lines arising from metastable levels around the visual spectrum}
\label{sec:2_1}
The gas shell surrounding the Beta Lyrae system was discovered long ago and its structure has always been studied on the basis of helium lines originating from metastable levels. First of all, it concerned the strongest emission-absorption line of the He\,I $\lambda$ 3888 in the violet spectrum. Well-known traditional studies of the broadening shell in the complex line of He\,I $\lambda$ 3888 \citep[e.g.][]{Struve1941,Sahade1959} were subsequently extended through the study of this shell in the infrared line of the He\,I $\lambda$ 10830 of this triplet series \citep[e.g.][]{Girniak1978}. It should draw attention that the contours of these lines contain narrow and deep absorption components with negative radial velocities in all orbital phases. If they exceed the parabolic velocities of moving microparticles in the binary system, it has traditionally been believed that the structural components of this shell, in which such lines are formed, move outward from this binary system. The He\,I $\lambda\lambda$ 3888, 10830 lines have a metastable level of $2^3$S. This level is highly populated with electrons and the effects of collision are crucial for the manifestation of these lines in the spectrum. Indeed, absorption components of these triplet lines of helium are well visible in the diluted shell structure around this binary system. At the same time, these helium lines have a very strong emission component in their contours, in many respects similar to those in lines H$_\alpha$  and He\,I $\lambda$ 7065 \citep[see][]{Skulskyy2020b}, where clear intercorrelations were discovered between the phase variability of the donor's magnetic field and dynamic and energy characteristics of these strong emission lines. This may indicate the general nature of the formation and spatial localization of all these emissions. It can be supposed that the absorption in the He\,I $\lambda\lambda$ 3888, 10830 lines, which cuts through their emission component, arises as self-absorption and should reflect the specificity of the matter loss from the donor surface, which, showing certain structural features of the magnetic field, reaches the Roche cavity. The advanced attempt to re-study the characteristic spectral features of these and other He\,I lines, which arise from metastable levels, in combination with the corresponding manifestation in the spectrum of the binary system of the donor magnetic field configuration presents new opportunities in this study.

\subsection{Magnetic field and He\,I lines arising from metastable levels in the violet spectrum}
\subsubsection{The above question and known article \mbox{by \cite{Sahade1959}}}
\begin{figure}[!t]
	\centerline{\includegraphics[width=0.9\textwidth,clip=]{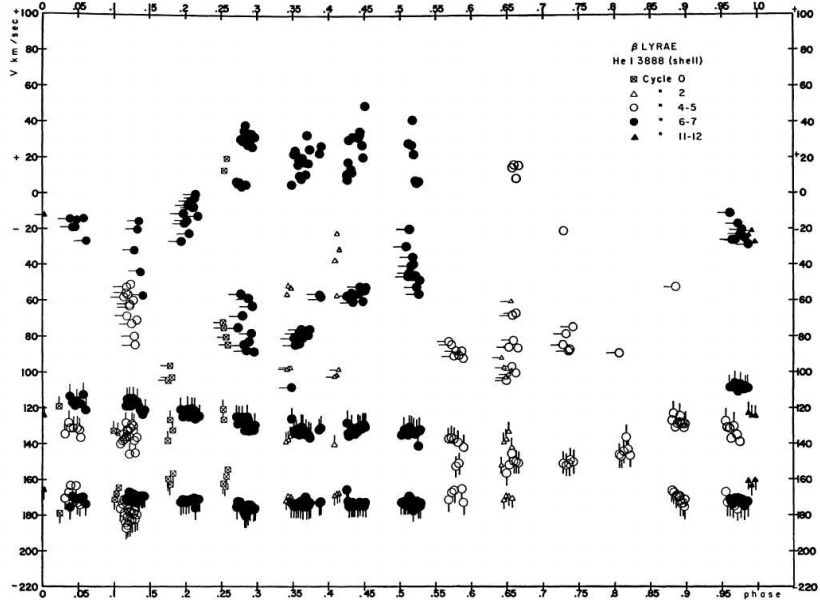}}
	\caption{Radial velocities from the shell (triplet) line at He\,I $\lambda$ 3888. The \textquote{tails} distinguish the groupings which the different components suggest. Two components yield rather constant velocities of about -170 and -130 km./sec. The rest of the components do not behave so regularly and show large differences in different cycles. Adopted from \cite{Sahade1959}.}
	\label{fig:1}
\end{figure}

Based on the concept of the existing configuration of the donor's magnetic field, the careful work of \cite{Sahade1959} can be considered as one of the most consistent and informative studies of the He\,I $\lambda$ 3888 line. The results of this investigation are based on studies of 195 plates with a dispersion of 10 \AA/mm obtained from observations of Beta Lyrae on the Mount Wilson 100-inch reflector over 12 orbital cycles in the 1955 season (the article presents an atlas of photographic spectrograms in the region $\lambda\lambda$ 3680-4580). However, only in 4-7 cycles in June-July they were performed most fully to be suitable for a detailed analysis. In Fig. \ref{fig:1}, which is a copy of Fig. 13 from the article of \cite{Sahade1959}, \textquote{radial velocity measurements from the shell line at He\,I $\lambda$ 3888} are presented. As it is seen at the bottom of Fig. \ref{fig:1}, where the two absorption components \textquote{with very definite} velocities of about -170 km/s and -130 km/s are shown, they converge to one absorption component near the 0.85\,P phase of the visibility of the magnetic pole on the donor surface. These absorption components demonstrate very clear sequences over the orbital phases, and, according to the understanding of \cite{Sahade1959}, their formation \textquote{must occur in a shell that surrounds the whole system}. Their radial velocities are the largest and differ in curves shape from such curves of radial velocities of the absorption components that are at the top of Fig. \ref{fig:1}. The latter more resemble the absorption components superimposed on emissions in lines H$_\alpha$ and He\,I $\lambda$ 7065 (reflecting the gas flows between the donor and the gainer; see Section 2.3 in \cite{Skulskyy2020b}).

\begin{figure}[!t]
	\centerline{\includegraphics[width=0.95\textwidth,clip=]{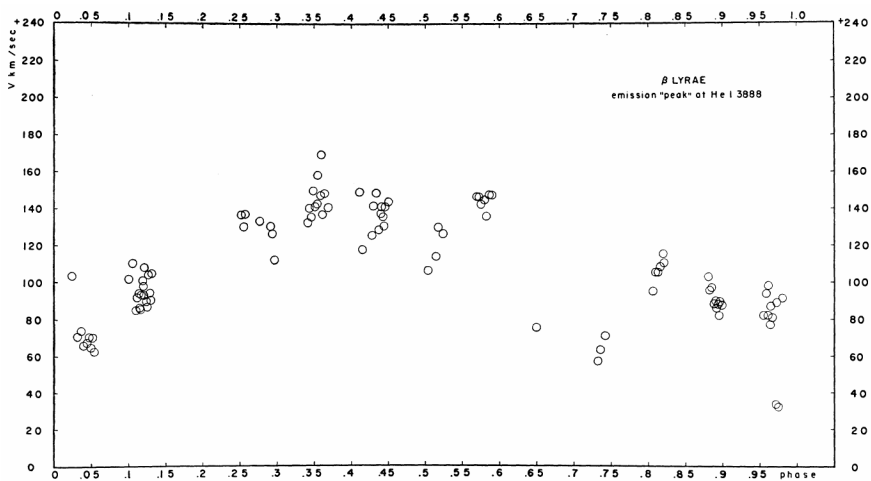}}
	\caption{Radial velocities from the emission "peak" at He\,I $\lambda$ 3888. Adopted from \cite{Sahade1959}.}
	\label{fig:2}
\end{figure}

It is well known that the He\,I $\lambda$ 3888 line also has an emission component, which consists only of the very high red peak in contrast to the two-peak emission, with superimposed absorption, at strong lines H$_\alpha$ and He\,I $\lambda$ 7065 (the latter are also above the continuous spectrum in all orbital phases). The violet peak in the He\,I $\lambda$ 3888 line is virtually absented owing to deep violet absorption, showing appreciable changes in observations of different investigators \citep[see, for example,][]{Skulsky2011}. Fig. \ref{fig:2}, which is a copy of Fig. 18 from the article of \cite{Sahade1959}, illustrates the radial velocities curve from the red emission peak at the He\,I $\lambda$ 3888 line. One sees two maxima at the 0.35\,P and 0.85\,P phases, i.e., in the phases of visibility of the magnetic field poles on the donor surface (the small local maximum in the 0.6\,P phase may reflect the collision of the gas flow with the accretion disk). The similar two maxima are observed on the radial velocity curve of the red peak at the He\,I $\lambda$ 4472 line (see Fig. 19 in \cite{Sahade1959}). The red peaks of these helium lines of the violet spectrum exhibit the same behavior as red peaks at the H$_\alpha$ and He\,I $\lambda$ 7065 lines. The spectral features of all components at the H$_\alpha$ and He\,I $\lambda$ 7065 lines were considered in detail on the basis of Figures 5 and 6 in \cite{Skulskyy2020b}. These maxima at the 0.35\,P and 0.85\,P phases match clearly the phases of the two maxima on the curve of the effective magnetic field strength of the donor \citep[see also Fig. 1 in][]{Skulskyy2020b}. Being synchronized in orbital phases, all these emission components have clear similarities, their physical nature and spatial interconnection seem indisputable. It can be assumed that the formation of red peaks in these lines, including such for the line He\,I $\lambda$ 3888, should be associated with the moving radiation near the donor surface, close to the poles of the magnetic field on this surface. This corresponds to the conclusions of \cite{Skulskyy2020b} on the study of dynamics of emission components in the H$_\alpha$ and He\,I $\lambda$ 7065 lines.

The dynamic behavior of both the above absorptions of the He\,I $\lambda$ 3888 line in \cite{Sahade1959}, which are shown at the bottom in Fig. \ref{fig:1}, is even more interesting. Their double nature has not been so clearly defined or repeated by investigators in other studies (see, for example \cite{Harmanec1993, Skulsky2011}, which analyzed such studies from the early 20th century). It should be noted that the radial velocity of the center of mass of this binary system, as determined from the interstellar Ca\,II lines by \mbox{\cite{Sahade1959}, is -15.2 km/s}. The $\gamma$ - velocity determined in Table 9, for the set of orbital elements of Beta Lyrae computed on the basis of all 192 spectral plates in the summer of 1955, is -16.0 km/s. The set of orbital elements, calculated from observations taken between 0.85\,P and 0.15\,P phases, i.e., with the data discarded during the main (or primary) eclipse, showed -15.0 km/s for the  -velocity. That is, both absorption components with radial velocities of approximately -170 and -130 km/s, should be considered as components with average velocities of about -155 and -115 km/s. Modern stellar masses of this binary system and the inclination of the orbit \citep[see in][]{Skulskyy2020a} allow affirming that the parabolic velocity of the moving gas particles reaches a value slightly smaller, but close to -115 km/s, which is sufficient to reach the Roche cavity with the Lagrange point L2. A simple calculation also shows \citep{Burnashev1991} 
that the velocity of the particles, which is sufficient to reach the Roche cavity with the Lagrange point L3, is near to -160 km/s. That is, the absorption component with an average radial velocity of approximately -155 km/s, or its possible greater values, may indicate a mass loss in a wider range of directions, in particular in the direction of the Lagrange points L4 and L5.

Here it is important to emphasize that \cite{Sahade1959} themselves noted: \textquote{there are striking changes with phase and cycle in the spectrum} and \textquote{the great complexity of the spectrum of Beta Lyrae makes it impossible to give a description which would do justice to the wealth of information contained in the material}. Indeed, at some points, this spectral material is presented as a general description. Now it is possible to consider the data of the observation in the light of modern ideas and concepts, in a more detailed manner. The behavior of its identified two absorption components of the He\,I $\lambda$ 3888 line should be investigated as carefully as possible, supplementing the analysis of radial velocities, given primarily in Table 3 from \cite{Sahade1959}. But here there are also several other lines of helium, in particular those arising from metastable levels, which on the basis of tabular data and graphs will help to significantly expand the scope of research. 

Based on Table 3, it is possible to estimate the behavior of the absorption of the He\,I $\lambda$ 3888 line with a smaller averaged radial velocity. In most orbital phases, Table 3 shows measurements of its radial velocities in the range of \mbox{125-135~km/s}. This corresponds to a matter outflow velocity of about -115 km/s relative to the center of gravity of the binary system, i.e., with the first parabolic velocity. The most \textquote{striking changes with phase and cycle in the spectrum} are visible in the phases (0.95-0.13)\,P within a wide primary eclipse (see Fig. \ref{fig:1}), when the donor is in the process of the eclipse by the gainer hidden in the accretion disk. At the time, according to the data of Table 3, the radial velocity of this absorption, during several days of observations at the boundary from 6 to 7 and from 11 to 12 orbital cycles, is observed in the range from -95 to -105~km/s, i.e., smaller than the parabolic velocity. The significant decrease in the radial velocity of this absorption component at the boundary of these orbital cycles indicates that the gas, flowing in the direction of the gainer in its Roche cavity, can only reach nearer structures of the accretion disk. But this is not observed in the same phases of the zero cycle and at the boundary of 4 to 5 orbital cycles, where the averaged radial velocity is approximately -115 km/s relative to the center of gravity. This indicates that the flows of matter during mass transfer are not laminar and have a variable nature within several cycles. Because in both events the motion of matter is visible above the accretion disk, i.e., the observer behind the gainer records a significant elevation of the moving matter above the plane of the orbit, this determines a considerable height of the rim of this disk from the side of the donor. As shown in \cite{Skulskyy2020b}, the vertical component of the gaseous flow in the direction of the gainer can be formed due to the configuration of the donor's magnetic field, in which the magnetic axis is directed at a significant angle to the plane of the orbit of this binary system.

It should be noted that in phases (0.95-0.13)\,P of the main eclipse, the more negative absorption component in the He\,I $\lambda$ 3888 line also shows the various radial velocities in different cycles. In phases (0.95-0.99)\,P on June 8 of the 4th cycle and on July 4 of the 6th cycle this absorption component shows the average radial velocity of -173 km/s (or -158 km/s relative to the center of gravity), reflecting the motion of matter over the accretion disk, which approaches the velocity of its exit outside the Roche cavity with the Lagrange point L3. The dynamics of gaseous flows recorded in phases (0.03-0.13)\,P of the 5th cycle when measuring the radial velocity of this absorption component on 8 spectrograms on June 9 and 16 spectrograms on June 10? is most intriguing. In phases (0.03-0.05)\,P of the first date, the average radial velocity was -170 km/s with its growth within 5 km/s, however, in phases (0.11-0.13)\,P on June 10, its average the velocity was -180 km/s (on the first 4 plates it was -175 km/s, on the next 4 plates in phase near 0.12\,P has reached the average velocity of -185 km/s, then it was -181 km/s and -180 km/s, respectively). Confident achievement of the radial velocity of -185 km (-170 km/s relative to the center of gravity) indicates that in this direction, the velocity of the particles was sufficient for the free escaping of the matter from the binary system with the intersection here of the boundary of the Roche cavity with the Lagrange point L3. But in the phases (0.03-0.05)\,P on July 5 and (0.12-0.14)\,P on July 6 of the 7th cycle, the average radial velocity was -171 km/s, which is much smaller than the velocity of -185 km/s in the 0.12\,P phase on June 10. Moreover, for example, in the phases (0.987-0.002)\,P at the boundary of cycles 11 and 12, this absorption component of the He\,I $\lambda$ 3888 line revealed a significantly lower average radial velocity of -163 km/s (or \mbox{only -148 km/s} relative to the center of gravity). That is, passing over the center of the accretion disk, the moving matter with such velocity, most likely, must be captured by the outer side of the revolving accretion disk. Therefore, the more negative absorption component of the He\,I $\lambda$ 3888 line also indicates that the matter flows during mass transfer are not laminar and have a variable character during one or several cycles.

\begin{figure}[!t]
	\centerline{\includegraphics[width=0.85\textwidth,clip=]{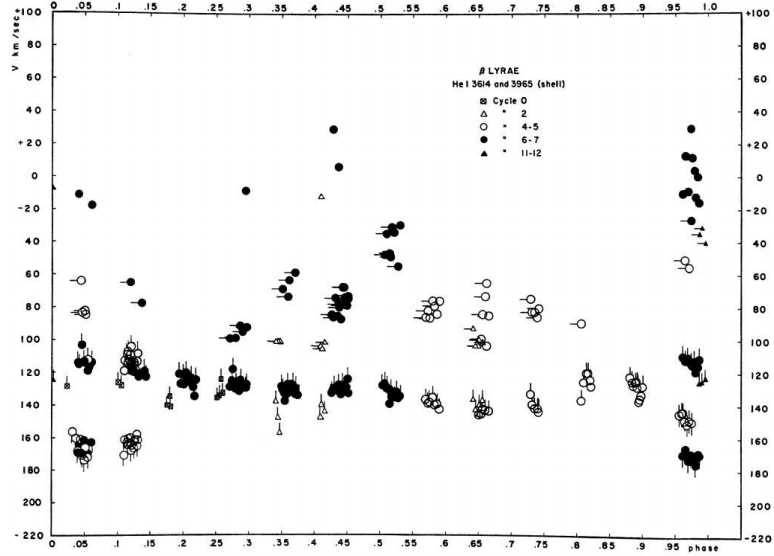}}
	\caption{Radial velocities from the shell (singlet) lines at He\,I $\lambda$ 3614 and 3964. The \textquote{tails} distinguish the groupings which the different components suggest. Adopted from \cite{Sahade1959}.}
	\label{fig:3}
\end{figure}

For comparison of the behavior of the He\,I $\lambda$ 3888 line during phases of the main eclipse, it makes sense to evaluate the behavior of other helium absorption lines arising from metastable levels. \cite{Sahade1959} measured the radial velocities from He\,I $\lambda\lambda$ 3614, 3965 lines of the singlet series. The radial velocities of their absorption components, as well as for the He\,I $\lambda$ 3888 line, are listed in Table 3 and shown in Fig.~\ref{fig:3} (as a copy of Fig. 14 from the article \cite{Sahade1959}). In almost all phases, the more negative component of these singlet lines forms a successive curve with an average radial velocity close to -115 km/s relative to the center of gravity (i.e., close to the first parabolic velocity) with complex deviations from it in phases (0.95-0.13)\,P. This also applies to phases (0.95-0.05)\,P around the middle of the main eclipse both for the zero's cycle (April 5 as the first observation date in 1955) and at the boundary of the 11th and 12th cycles (September 7 as the last observation date). But an important difference in the variability of these lines around the middle of the main eclipse concerns the observation periods of June 8-10 at the boundary of the 4th and 5th cycles and July 4-5 at the boundary of the 6th and 7th cycles. It is in these June and July observations that two narrower absorption features in these singlet helium lines are clearly distinguished. The development of events should be considered from the end of the 4th cycle, noting that on June 7 in the phases (0.88-0.90)\,P of direct visibility of the magnetic pole on the donor surface there was the average radial velocity of this absorption of about -115 km/s relative to the center of gravity. A clear rise of the radial velocity of the above absorption at its increment from -130 km/s to -135 km/s for 8 spectrograms occurred on June 8 within the phases (0.96-0.98)\,P. On June 9, within the phases (0.03-0.05)\,P, this absorption component during 8 spectrograms reached the radial velocity from -146 km/s for the first 4 spectrograms to -156 km/s for the next 4. Such velocity of moving matter is close to the second parabolic velocity. However, on June 10, within the phases (0.11-0.13)\,P, this average radial velocity on all 16 spectrograms fell slightly to -148 km/s with respect to the center of gravity. For the first time, these spectrograms clearly manifested the second component of absorption that had an average radial velocity of \mbox{-97~km/s}, somewhat smaller than the parabolic velocity (near -115 km/s relative to the center of gravity). On July 4 of the 6th cycle, 8 spectrograms within the phases (0.96-0.99)\,P recorded an average radial velocity of -155 km/s for the first absorption component and -98 km/s for the secondary one. On July 5, 7 spectrograms within phases (0.04-0.06)\,P of the 7th cycle recorded an average radial velocity of -151 km/s for the first absorption component and -98 km/s for the secondary one. However, the next night, July 6, unexpected changes were recorded: all 12 spectrograms in phases of (0.12-0.14)\,P showed that the radial velocity of the first absorption component dropped to -105 km/s, and the second absorption component disappeared regarding the measurement of radial velocities.

Thus, in the June and July observations, these singlet helium lines, becoming double in certain phases of the main eclipse, reveal asynchronous differences. Between the two orbital rotations from June 10 to July 6, in the same phases (0.11-0.14)\,P, the spectrum of the binary system represents very different expressed sets of both absorption components of these lines. If on June 10 both components of these lines became clearly double at radial velocities of -148 km/s and -98 km/s, then after two orbital cycles on July 6 the first of the two absorption components diminished its velocity to -105 km/s, i.e., slightly less than the parabolic velocity, but the second absorption component was absent. Moreover, if in the 5th cycle in June both components of these lines with velocities of -148~km/s and -98~km/s became visible only in the phases (0.11-0.13)\,P, then in the 6th cycle similar radial velocities -155~km/s and -98~km/s of both these components were measured in phases (0.98-0.99)\,P on July 4 and close to such velocities -151 km/s and -98 km/s on July 5 (in the 7th cycle) in phases (0.04-0.06)\,P (the next night on July 6, in phases (0.11-0.14)\,P the first absorption component had the radial velocity only -105~km/s, and the second absorption component unexpectedly disappeared and was invisible on the following nights of this 7th cycle). Hence, it should be noted that within the main eclipse the He\,I $\lambda\lambda$ 3614, 3964 lines show two phase segments of significant changes in their intensity and dynamics, which may indicate disordered changes in the density of the accretion disk. They are grouped close to the 0.95\,P and 0.05\,P phases, which, reflecting the formed dynamically changing structures of the disk, correspond simultaneously to the phases of the maximum manifestation of satellite lines (see \cite{Skulskyy2020a} and more fully \cite{Skulskij1992}). The third region of such changes is formed by the denser matter flow in the phases (0.10-013)\,P, which opens after the center of the eclipse. 

Comparing the changes in the radial velocities in the absorption components of singlet and triplet helium lines arising from metastable levels, it should be especially noted the observation on June 10 in the phase near 0.12\,P. In the He\,I $\lambda$ 3888 line, there was recorded the radial velocity -170 km/s relative to the center of gravity, which is sufficient for the motion matter from the binary system in any direction, including its exit outside the Roche cavity with the Lagrange point L3. At the same time, for the first time, a clear bifurcation of He\,I $\lambda\lambda$ 3614, 3964 lines (of the singlet series) into two absorption components with radial velocities comparable to two parabolic velocities was recorded. It is also important that Table~4 and Fig.~15 in \cite{Sahade1959} show that these observations on June 10 recorded on all 16 spectrograms the strongly shifted radial velocities for absorptions, which cut through emissions of H$_\gamma$ and \mbox{He\,I $\lambda\lambda$ 4472, 4026} lines. Shifts in the position of these absorptions make up -151 km/s (as the average on the first eight spectrograms) and -141 km/s (on the last eight of them) relative to the center of gravity of the binary system. A similar phenomenon was recorded only during the zero's cycle on April 6 in phase 0.10\,P, when measurements of these lines showed on the obtained spectrograms the radial velocity of -157 km/s relative to the center of gravity of the binary system, at clear excess of the second parabolic velocity, sufficient to an outflow matter via stellar wind from the Lagrange point L3.

The events described above, recorded in the behavior of the radial velocities of the absorption components of helium and hydrogen lines in the violet spectrum, show, within the main eclipse, not laminar but to some extent an unpredictable sharply variable mass transfer between the components of the binary system. In close phases of different dates, the observer behind the mass gainer registers the approach of the matter from the donor to the gainer at different radial velocities above the surface of the accretion disk. It is clear that when such velocities of a matter are slightly smaller than the first parabolic velocity, this matter settles down on the surface of the accretion disk, from the donor side. When the radial velocities of the moving substance exceed the first parabolic velocity or are a significant fraction of the second parabolic velocity, such gas flows can be captured directly by the gainer or the outer parts of the accretion disk. Repeatedly recorded velocities of matter surpassing the second parabolic velocity confirm the direct exit of matter beyond the close circular structures of the binary system and the possibility of forming a distant shell. This agrees with the observations of the radio-nebula, which surrounds the Beta Lyrae system, indicating the substantial non-conservative mass-loss during the evolution of this interacting system \citep{Umana2000}. The obtained generalized picture does not yet have data that would allow us to make specific predictions of the above events. They can be caused by a holistic system of secondary periods and resonances that change synchronously with the growth of the orbital period, variable substructures of the outer layers of the accretion disk, the peculiar spatial configuration of the donor's magnetic field, or other reasons considered in \cite{Skulskyy2020a, Skulskyy2020b}. In addition, the findings on the variable dynamic behavior of the absorption components of the above helium lines may to some extent confirm the results of disk-forming motions of matter in gas-dynamic modeling of stellar structures and accretion disk, given by \cite{Bisikalo2000}.

Analyzing the behavior of the radial velocities of the absorption components in complex helium and hydrogen lines, the previous points in this section focused on the details of the transfer of matter from the donor to the gainer within the main eclipse, i.e., in the direction of the gravitational axis of the binary system. It is equally important to evaluate their behavior in the phases associated with the transfer of moving matter in the direction of the axis of the donor's magnetic field. It should be noted \citep{Skulskyy2020a} that a dipole magnetic field with the axis in the direction of the orbital phases (0.355-0.855)\,P has a maximum magnetic field on the donor surface at the 0.855P phase, which reflects the location of this magnetic pole close to the gainer. In addition, the magnetic dipole axis inclined to the orbit plane of the binary system by 28\degr, and therefore, this magnetic pole on the surface of the donor in phase 0.855P is also above the plane of the orbit. This means that the ionized gas directed by the donor's magnetic field moves in the direction of the dipole axis from the donor's surface and, deviating along the magnetic field lines towards the accretion disk, can rise above the front edge of the disk without much energy loss. 

The tables and graphs in \cite{Sahade1959} allow of stand out certain aspects related to the moving matter in the direction of the axis of the magnetic field of the donor. It is reasonable to estimate first the behavior of radial velocities of absorption components of the above lines in orbital phases near 0.85\,P, i.e. in the phases of direct visibility on the donor surface of the magnetic pole facing the gainer. From Fig.~\ref{fig:1} it is seen that in phases around 0.8\,P both absorption components in the He\,I $\lambda$ 3888 line form the single absorption with the radial velocity near -135~km/s. This is sufficient for a direct entering of matter in to an external medium in this direction through the Roche cavity with the Lagrange point L2 (the first parabolic velocity relative to the center of gravity of the binary system is near -115~km/s). In these phases, the more negative absorption component of the He\,I $\lambda\lambda$ 3614, 3964 singlet lines shows a single average radial velocity close to -115~km/s (not showing any bifurcation, see Fig.~\ref{fig:3}). Several absorption groups with much smaller radial velocities, which are present in other phases, are completely absent in phases before 0.855\,P, i.e., the flow of magnetized plasma is manifested as a single whole. In the phases (0.88-0.90)\,P after 0.855\,P, the physical conditions changed. The more negative absorption component in the He\,I $\lambda$ 3888 line reached the average radial velocity of -156~km/s, and the absorption of the He\,I $\lambda$ 3888 line with a lower average radial velocity approached to a radial velocity near -115~km/s, i.e., to the second and first parabolic velocities, respectively. Evidencing the certain stratification of their formation, the He\,I lines of both series arising from metastable levels emphasize the average direction of plasma flows in the direction of the donor's magnetic field axis and localization in phases around 0.855\,P on the donor surface of the magnetic pole close to the gainer. This indicates that in the phases of direct visibility on the donor surface of this magnetic pole there is an outflow of magnetized matter outwards from the donor surface, as one of the important directions of non-conservative loss of matter during the evolution of this interacting system.

Based on \cite{Sahade1959} observations in 1955 it can be stated that the radial velocities in the phases near both poles of the magnetic field in the He\,I $\lambda$ 3888 absorption line showed a radial velocity much higher than the first parabolic velocity (see Fig. \ref{fig:1}). However, the stellar wind behaves somewhat differently in the phases of the first quadrature, in particular in the vicinity of the 0.355P phase of the magnetic field pole on the donor surface. This applies, for example, to both the phase interval (0.25-0.5)\,P in the He\,I $\lambda$ 3888 line (see Fig.~\ref{fig:1}) and phases of (0.25-0.45)\,P in the H$_\alpha$  and He\,I $\lambda\lambda$ 4472, 4026 lines (see Figures 13, 15, and 16 and Tables 3 and 4 in \cite{Sahade1959}). Several groups of radial velocities inherent to the corresponding absorption components are visible. The group of absorptions with positive radial velocities around +20 km/s, i.e., +35 km/s relative to the center of gravity of the binary system is important. It indicates that within the (0.35$\pm$ 0.1)\,P phases of the visibility of the magnetic pole on the donor surface, a runoff of magnetized matter from the observer is formed in the direction of the axis of the donor's magnetic field (0.35-0.85)\,P. Reflecting the thermal velocity of matter from the surface of the donor, which reaches its Roche cavity, the radial velocity in these phases indicates the direction of the stellar wind from the surface of the donor. It is directed from the observer along the magnetic axis of the donor to the visibility phases of 0.855P on the donor surface of the magnetic pole facing the gainer, where the magnetic field of the donor reaches its maximum value. As it can be seen from Fig.~\ref{fig:3}, such a group of radial velocities is not clearly detected in the lines He\,I $\lambda\lambda$ 3614, 3964, indicating some stratification of physical conditions in the moving matter. But these lines, like the lines He\,I $\lambda$ 3888, H$_\alpha$ and He\,I $\lambda\lambda$ 4472, 4026 have absorption components, which clearly are changed in radial velocities from -60 km/s to -40 km/s in phases of (0.30-0.45)\,P. This exhibits certain variable events near this pole of the donor's magnetic field that are related, for example, to variations in these phase ranges of the total flux of polarization recorded in \cite{Hoffman1998} and \cite{Lomax2012}. To some extent, these changes were studied on the basis of helium lines arising from metastable levels by \cite{Skulsky2011}. Now, this requires detailed coverage and comparison with the evidence of \cite{Sahade1959}. 

\subsubsection{Helium lines arising from metastable levels in more modern studies}
\begin{figure}[!t]
	\centerline{\includegraphics[width=0.75\textwidth,clip=]{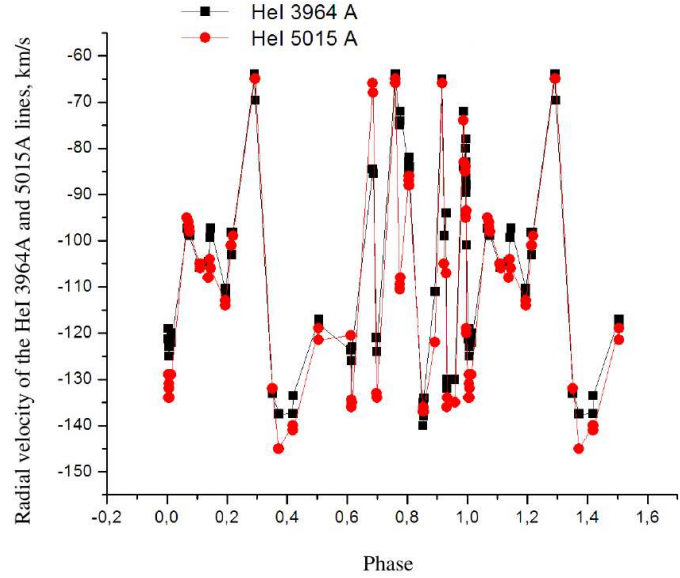}}
	\caption{Variations in the radial velocities curve of the He\,I $\lambda\lambda$  3964, 5015 lines. Adopted from \cite{Skulsky2011}.}
	\label{fig:4}
\end{figure}

As it was previously pointed out, the issues of the dynamics of circumstellar gaseous structures using lines He\,I $\lambda\lambda$  3888, 3964, 5015, which arise from metastable levels, were studied by \cite{Skulsky2011}. During 2008-2010, they obtained almost 100 echelle spectrograms with a resolution of about 45 000 on the 2-m reflector of the Peak Terskol Observatory. The narrow absorption components of the He\,I $\lambda\lambda$ 3964, 5015 showed that the variability of radial velocities and intensities with the phase of the orbital period is practically correlated: both lines formed at approximately the same distance from the center of the system and participate in the same processes. The intensities of these lines reach a minimum in the phases near (0.45-0.60)\,P, when the donor closes the gap between it and the gainer, simultaneously highlighting the phase direction (0.6-0.1)\,P perpendicular to the axis of the donor's magnetic field. The intensity of these lines also decreases sharply within less than 0.1\,P in the phases of 0.35\,P and 0.85\,P of polar regions on the donor's surface (more obviously in the He\,I $\lambda$ 3964 line, see Fig. 1 in \cite{Skulsky2011}). But particularly sharp changes are clearly seen in the negative radial velocities of the He\,I $\lambda\lambda$ 3964, 5015 lines, which indicates complex processes in the orbital structures of this binary system (see Fig. \ref{fig:4}, which is a copy of Fig. 2 from \cite{Skulsky2011}). It can be seen that within 0.1\,P in the 0.35\,P and 0.85\,P phases, which correspond to the location of the magnetic poles on the donor surface, there is a sharp drop in the radial velocity from -55 km/s to -125 km/s, i.e., by 70 km/s relative to the center of gravity of the binary system ($\gamma$ = -18 km/s according to measurements of the radial velocities of the sharp interstellar line Ca\,II $\lambda$ 3933). 

The behavior of the radial velocities curve of these helium lines is actually clearly associated with the magnetic field curve obtained in the course of measurements of the Zeeman splitting at Si\,II $\lambda\lambda$ 6347, 6371 lines by \cite{Skulskij1993}. The effective intensity of the longitudinal component of the magnetic field varies during the orbital period within $\pm$ 200\,G, but as it can be seen from Figures 5 and 6 of \cite{Skulskyy2020a}, rapid changes in the magnetic field curve are clearly visible only in the phases near the poles of the magnetic field. In both cases, the polarity of the magnetic field varies within 0.1\,P in the 0.355\,P phase of the first quadrature and in the 0.855\,P phase of the second quadrature. The behavior of the photographic curve of the magnetic field is also characterized by rapid changes in the magnetic field curve around the 0.855\,P phase (i.e. the phase of observation of the magnetic pole facing the gainer) and in the phases around the main eclipse.

Note that the sharp peak of radial velocities with a rapid change in their values is observed in Fig. \ref{fig:4} in the phase of about 0.67\,P (one can assume that the opening gas flow is in collision with the accretion disk). Similarly formed two narrow peaks of radial velocities are observed in the same phases of the main eclipse, in which the radial velocities of the opposite sign are observed in the satellite lines, characterizing the outer edges of the accretion disk. It should be concluded that due to different physical parameters at the edges of the accretion disk a complex structure of plasma flows is formed, which may participate primarily in mass transfer between components of the binary system, as well as in the outflow of matter outside it. In addition, note that the rapid changes of radial velocities near the main eclipse in the He\,I $\lambda\lambda$ 3965, 5015 lines (see Fig.~\ref{fig:4}) are to some extent related to such events recorded in the similar phases (0.88-0.90)\,P and (0.96-0.98)\,P of radial velocities from the He\,I $\lambda\lambda$ 3614, 3965 lines according to their measurements in \cite{Sahade1959} (their behavior was discussed above since the end of the 4th cycle on June 8, 1955). Such a coincidence is not accidental. 

\begin{figure}[!t]
	\centerline{\includegraphics[width=0.85\textwidth,clip=]{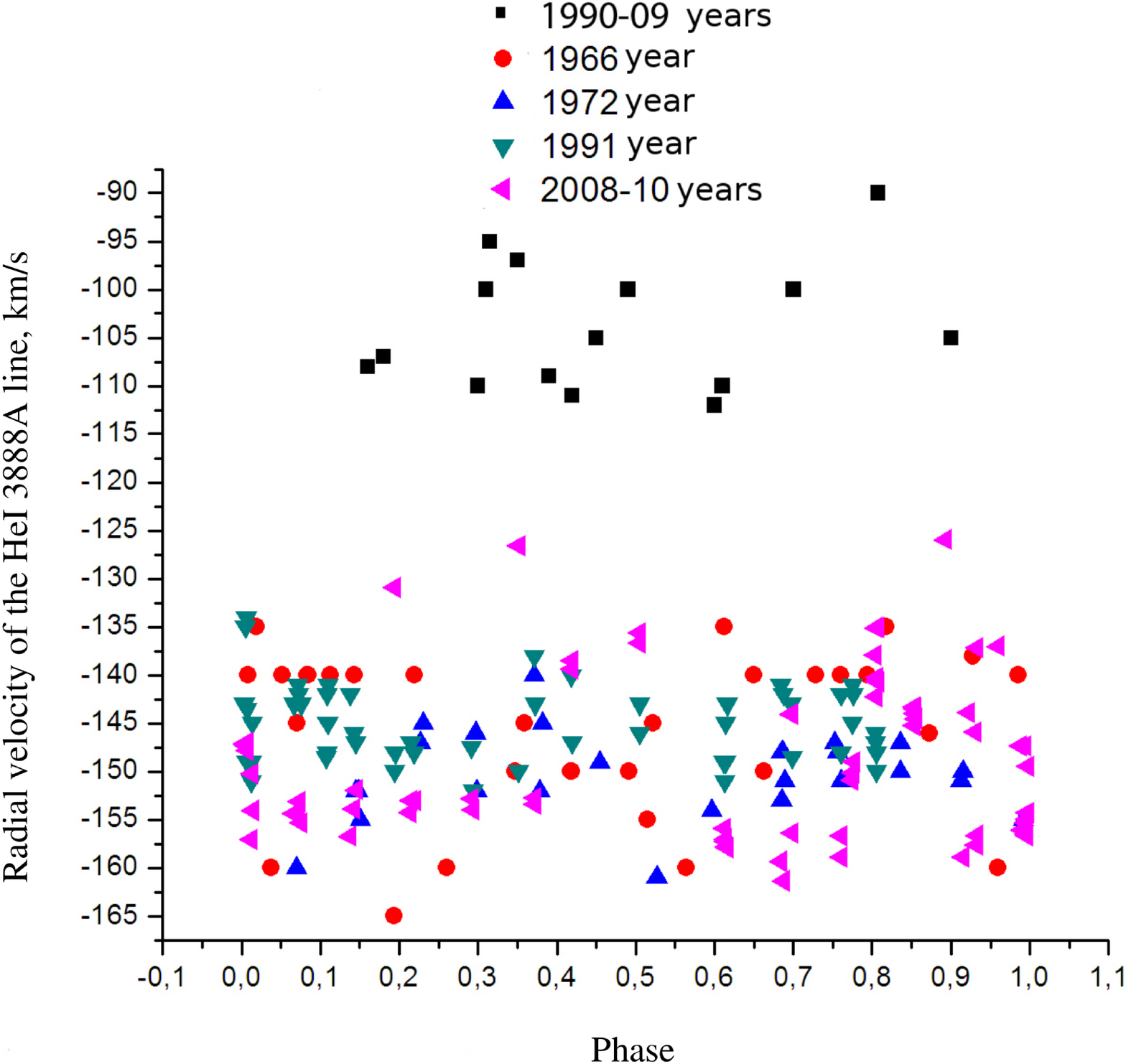}}
	\caption{The radial velocity of the He\,I $\lambda$ 3888 line in 1900 - 2010. Adopted from \cite{Skulsky2011}.}
	\label{fig:5}
\end{figure}

Based on observations of the Peak Terskol Observatory in 2008-2010, \cite{Skulsky2011} also investigated the behavior of radial velocities of the He\,I $\lambda$ 3888 line for 1900-2010. Some dynamic characteristics are shown in Fig. \ref{fig:5} as a copy of Fig. 3 from their paper. Without commenting on the nature of the peculiar understatement of radial velocities on the Potsdam 1900-1909 spectra (one can agree with the thorough analysis of this problem conducted by \cite{Harmanec1993}), one should note certain points (given the average radial velocity of the center of gravity of this binary system $\gamma$ = -18 km/s, which is determined within 2 km/s on the spectrographs of the Crimean, Ondrejov and Peak Terskol Observatories). The range of measurements of radial velocities according to our Crimean observations in 1966-68 and 1972, our more recent observations in 2008-2010 at Peak Terskol, as well as observations in 1991 in  Ondrejov, lies in a range from -118 km/s to -142 km/s at the average radial velocity near -130 km/s. There are no such sharp jumps in radial velocities inherent in the curve of radial velocities of the He\,I $\lambda\lambda$ 3964, 5015 sharp lines shown above in Fig. \ref{fig:4}. Well-provided observations of the 1991 season at the Ondrejov Observatory show the average radial velocity in the He\,I $\lambda$ 3888 line, which is only about -127 km/s, but the radial velocity curve based on the data in Table 4 and Fig. 3 by \cite{Harmanec1993} shows a hump with a value of -122 km/s in the phase range (0.70-0.85)\,P of the visibility of the magnetic pole on the donor surface (and -132 km/s in the phase of 0.05\,P). This is consistent with the convergence of both absorption dependences on He\,I $\lambda$ 3888 (see Fig. \ref{fig:1}) to a single absorption component, which yields a reduced radial velocity to -(130-135) km/s in the phase range (0.75-0.85)\,P of the visibility of this magnetic pole on the donor surface as shown in \cite{Sahade1959}. In both  papers, the flow of matter in the direction of the axis of the donor's magnetic field is significantly greater than the first parabolic velocity. If the difference of 10 km/s in radial velocities between two distant seasons of active observation is real, then this could be a confirmation of long-term changes in the non-laminar loss of moving matter out of the circumstellar structures of this binary system. It should be noted that none of the researchers of the He\,I $\lambda$ 3888 line observed this type of bifurcation of its absorption into two different components during the almost  entire orbital period. In this aspect, the article by \cite{Sahade1959} remains original and its consideration required special research. This is one of the reasons why this issue was not considered in the report of \cite{Skulsky2011}. Note that the He\,I $\lambda$ 3888 line behaves somewhat differently in the known article by \cite{Flora1975}. This thorough article also deserves a detailed review, which for some reason has not been conducted for a long time. 

\subsubsection{Magnetic field and behavior of shell lines in the known article by \cite{Flora1975}}
In this study, it is necessary to consider the radial velocity curve of the He\,I $\lambda$ 3888 sharp absorption line in the paper of \cite{Flora1975}, the variability of which seems somewhat unusual. In most orbital phases these radial velocities should now be considered much smaller than the first parabolic velocity for this binary system (perhaps this is one of the reasons for the inactive use of this work). \cite{Flora1975} gave the results of the radial velocity (RV) measurements and the profiles and intensities of the most significant lines, which were taken during the international campaign in the period of July 18 - August 1, 1971. The rich spectrographic material (lacking only one day at the orbital period that is about 13 days) consisted of 70 spectrograms of the enough high-dispersion (7 and 12 \AA/mm) obtained at the 1.52-m telescope of the Haute Provance Observatory. \cite{Flora1975} indicate \textquote{an asymmetrical distribution of matter in the envelope surrounding the whole system}. They conclude that the behavior of the RV curves for the Balmer and He\,I shell absorption cores must be real because these lines are virtually independent of blending with absorption lines of the B9 star (or now the donor). It is also noted that \textquote{none various absorption or emission features show any correlation with the invisible companion} (or now the gainer). This motivates to consider  their rich observational material, drawing on modern concepts in the approach to its interpretation. 

\begin{figure}[!t]
	\centerline{\includegraphics[width=0.85\textwidth,clip=]{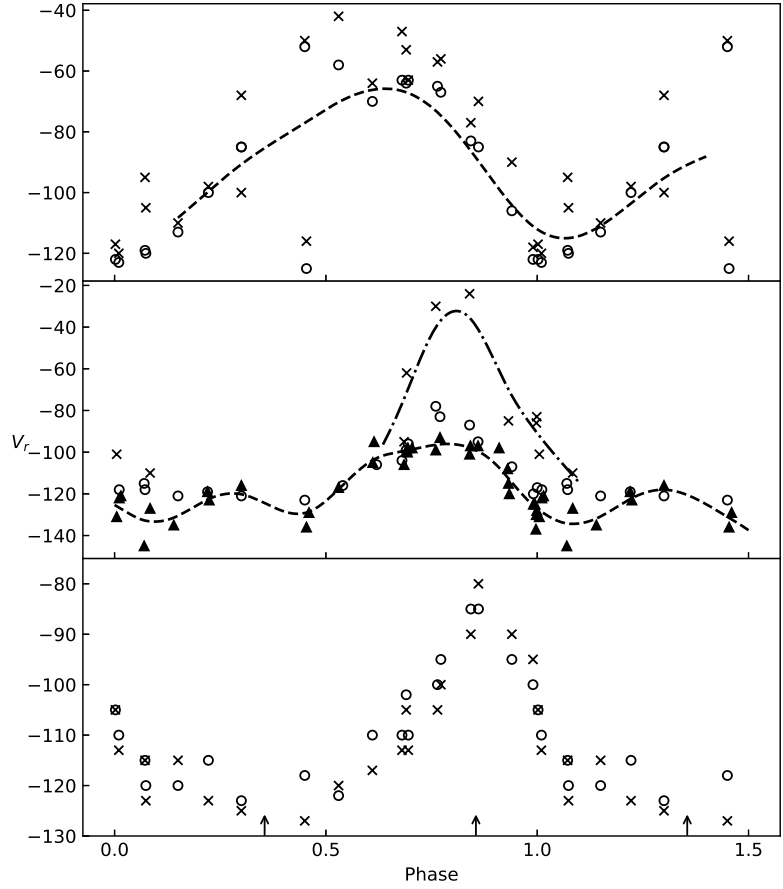}}
	\caption{The radial velocities (plotted according to data of \cite{Flora1975}) for certain shell lines in the Beta Lyrae spectrum.
 At the top there are lines He\,I $\lambda$ 5875 (circles) and He\,I $\lambda$ 6678 (crosses); in the middle, there are lines He\,I $\lambda$ 3888 (triangles), He\,I $\lambda$ 5015 (circles) and He\,I $\lambda$ 3447 (crosses); at the bottom there are lines Na\,I $\lambda$ 5889 (crosses) and Na\,I $\lambda$ 5895 (circles). The arrows in the 0.355\,P and 0.855\,P phases indicate the direction of the poles of the donor's magnetic field.}
	\label{fig:6}
\end{figure}

Fig.~\ref{fig:6}, along with other spectral lines, shows the data of measurements of the radial velocities of the He\,I $\lambda$ 3888 line, taken directly from the tables of the article of \cite{Flora1975}. If the average velocity of the center of gravity of the binary system is -17 km/s (according to Table 3 in interstellar components line Ca\,II $\lambda$ 3933 it is -15 km/s, and in lines of Na\,I $\lambda$ 5899 and 5895 it is -19 km/s), all applied radial velocities should in Fig. \ref{fig:6} shift up 17 km/s, in the direction of reducing negative velocities. Then, in phases (0.65-0.85)\,P, the He\,I $\lambda$ 3888 line will show the radial velocities of about -80 km/s relative to the center of gravity of the binary system (which is actually much smaller than the current first parabolic velocity for this binary system). This result immediately revealed several important points in relation to further research. First, the radial velocity -80 km/s is comparable to the average radial velocity obtained on Potsdam spectrograms in the early 20th century, the reality of which has been repeatedly discussed and questioned (see, for example, \cite{Harmanec1993}). Second, the stability in profiles, intensity, and radial velocities of the He\,I $\lambda$ 3888 line in orbital cycles during the 1971 season can be questioned, based, for example, on significant changes in these spectral characteristics in the observations of \cite{Sahade1959}. The question arises of changing the directions of active loss of matter both between stellar components and in relation to circumstellar medium, as well as the reasons for the variable density of such outflows.  

Regarding the direction of interpretation of Fig.~\ref{fig:6}, it is necessary to pay attention to its three parts, which, according to the classification of \cite{Flora1975}, present the radial velocity curves of two of the non-metastable He\,I $\lambda\lambda$ 5875, 6678 lines of the highest intensities, the metastable lines of He\,I $\lambda\lambda$  3888, 5015, 3447, and the shell components of the Na\,I $\lambda\lambda$ 5889, 5895 lines. One can consider a central part of Fig.~\ref{fig:6}, which shows three of the five lines measured in \cite{Flora1975} arising from metastable levels. The absorption components of these lines form dome-shaped curves of radial velocities in the phases of the second quadrature. The maxima of these dome-shaped curves were recorded in consecutive days of observations in the phases (0.78 - 0.77)\,P and (0.84 - 0.86)\,P. Similar curves show two more such lines He\,I $\lambda$ 3964 and He\,I $\lambda$ 6313, the radial velocities of which in these phases are located mainly between the curves of radial velocities of the He\,I $\lambda$ 5015 and He\,I $\lambda$ 3447 lines (in Fig. \ref{fig:6} there were not introduced so as not to overload the general picture). All five lines of these helium lines with certain potential difference between the upper and lower levels demonstrate clear stratification in the radial velocities, but the matter outflows have velocity much smaller than the parabolic velocity. 

Since the observer in phases (0.75 - 0.85)\,P looks perpendicularly to the line of centers of stellar components and along the axis of the dipole magnetic field of the donor (the motion of matter to the observer), the most adequate explanation of this picture can be considered as the consequences of loss of matter directly from the donor surface near its magnetic pole in phase 0.855\,P and its subsequent transfer to the gainer. It should be noted that the He\,I $\lambda$ 3888 line shows the single absorption, the radial velocity of which confidently reaches the current value of the first parabolic velocity only in the vicinity of phases (0.0 - 0.1)\,P, i.e. in the directions of motion of the matter to the gainer, as well as in phases near 0.45\,P. It is known that in phases (0.42 - 0.49)\,P a hot spot is registered on the accretion disk, according to \cite{Burnashev1991} and \cite{Lomax2012}. Here, the interpretation of such velocities can be related to the outflow of matter from the donor surface or, conversely, to this surface, reflecting the specifics of the location of the pole of the magnetic field on this surface. Indeed, \cite{Burnashev1991} showed that the observer begins to see excess radiation (hot surface on the donor) only from the 0.355P phase, i.e., along the axis of the magnetic field when its pole opens on the donor surface (see Fig. 1 in \cite{Skulskyy2020b}). This is more likely when the pole of the magnetic field is significantly shifted to the upper hemisphere on the surface of the donor (the magnetic dipole axis is inclined to the orbit plane of the binary system by 28\degr~ \citep{Skulskij1985}. Indeed, the hot surface on the donor continues to be observed almost to the 0.50P phase, demonstrating during the orbital phases (0.38-0.49)\,P the clear variability of the absolute radiation flux in the H$_\alpha$ emission line and the rapid variability of the spectrum in the H$_\alpha$ emission region \citep[see also][]{Alekseev1989}. This may be a reflection of the collisions of the hot plasma with the donor surface, directed (to the observer with the radial velocity close to the first parabolic velocity) along the lines of force of the magnetic field. Note that in these phases, there is a bifurcation of the absorption component so that both shallow absorption components are superimposed on the strong emission component of lines H$_\alpha$, He\,I $\lambda$ 6678, and He\,I $\lambda$5875, respectively, in Fig. 3d, 6a, and 7a in \cite{Flora1975}. The more violet-shifted component of this absorption shows that the tabular radial velocity approaches in the phase of 0.45\,P to current values of parabolic velocity (see Fig. \ref{fig:6}). Their occurrence can also be explained by the motion of the magnetic plasma along the lines of force of the magnetic field. In the He\,I $\lambda\lambda$ 5875, 6678 lines the red-shifted component of this absorption shows much smaller radial velocities close to the velocities in the lines of the donor atmosphere. Interestingly, \cite{Flora1975} suggested that the violet-shifted absorption in phase 0.45\,P may reflect the escape of gas matter through the Lagrange point in front of the donor. Such a hypothesis requires more observational evidence.

The radial velocity curve of the sodium lines resonates with the previous remark. The bottom of Fig.~\ref{fig:6} does not so clearly show that the observation phase of 0.375\,P (near the phase of 0.355\,P of direct vision of the pole of the magnetic field on the surface of the donor) was missed. In the phase range (0.30 - 0.45)\,P one can assume both a certain minimum of this curve near the 0.355\,P phase and a small local hump. In each case, the curve of radial velocities of the Na\,I $\lambda\lambda$ 5889, 5895 absorption lines clearly manifests itself as the wide maximum of negative velocities near the 0.355\,P phase and their clear narrow minimum in the phase 0.855\,P, i.e., in the phases of visibility of the magnetic field poles according to the photographic curve of the effective magnetic field strength of the donor (see Fig.~1 in \cite{Skulskyy2020b}). The presence of narrow interstellar components of the Na\,I lines allowed us to check the tabular data and to clarify the radial velocities of the shell components of these sodium lines directly from Figs. 7a and 7b of \cite{Flora1975}. Fig.~\ref{fig:6} shows that the observation in phases (0.8 - 0.9)\,P is especially interesting. Since the interstellar components in Na\,I $\lambda\lambda$ 5889, 5895 lines are -19 km/s, the radial velocity in the direction of the observer in these phases is the smallest one and is only of about -70 km/s, which is much smaller than the first parabolic velocity. Here the observer looks perpendicular to the donor surface along the axis of the magnetic field. Important is the fact of rapid dynamic changes in the moving plasma in the ranges of 0.1\,P in phases near 0.855\,P, i.e., of the center of the magnetic pole on the donor surface. Within 0.1\,P, the movement of the plasma clearly slows down, stops at a certain level, and accelerates again. Hence, the observations of \cite{Flora1975} clearly state the fact of actual variations in the sign of the RV progression curves of both helium and sodium lines reflecting the direction of motion of matter around the 0.855\,P phase. It can be argued that the observations of \cite{Flora1975} during the 1971 international campaign clearly identified the region of the donor surface that is responsible for the location of the pole of the magnetic field facing the gainer. This is consistent with studies of \cite{Zeilik1982} and \cite{Skulsky2011}, confirming CCD research of the magnetic field by \cite{Skulskij1993}.

Radial velocities in the Na\,I $\lambda\lambda$ 5889, 5895 lines in Fig.~\ref{fig:6} do not show a detailed picture of their possible variability in phases (0.95 - 0.05)\,P of eclipsing of the donor by the gainer wrapped in the accretion disk. The observer sees in phases (0.60 - 0.85)\,P a gradual decrease in the radial velocities by 30 km/s, which in phase 0.855\,P immediately changed to their such gradual increase in phases (0.85 - 0.10)\,P. One should think that the spatial configuration of the donor magnetic field reflects the direction (0.6 - 0.1)\,P, perpendicular to the direction (0.35 - 0.85)\,P of the donor magnetic axis. This is the direction of movement of the plasma perpendicular to the lines of force of the magnetic field in the space between the stellar components under the action of gravity of a massive gainer. 
It should also be noted that the radial velocity curves of the He\,I $\lambda\lambda$ 5875, 6678 absorption lines show in certain respects similar behavior. 
Their average curve at the top in Fig. \ref{fig:6} is made using measurements of the radial velocities of the line He\,I $\lambda\lambda$ 5875 as a line, the theoretical intensity of which is seven times greater than that of the He\,I $\lambda$ 6678 line. 
The radial velocities of the He\,I $\lambda$ 6678 line are located slightly higher than the velocities of the He\,I $\lambda$ 5875 lines, but in general, the behavior of these lines with the orbital phase is similar. 
The extreme values of their radial velocity curves correspond to phases of approximately 0.6\,P and 0.1\,P, reflecting the direction (0.6 - 0.1)\,P, perpendicular to the magnetic axis of the donor with the direction (0.35 - 0.85)\,P. 
These curves are shifted relative to the 0.855\,P phase by about a quarter of the orbital period, as it is seen in comparison, for example, with the RV-curves in Fig.~\ref{fig:6} for the Na\,I $\lambda\lambda$ 5889, 5895 lines. 
The absorption component with radial velocities, which changes rapidly in the phases near (0.60 - 0.65)\,P, is well seen in the higher members of the singlet lines of helium, in particular, as shown in Fig.~\ref{fig:6} in the He\,I $\lambda$ 3447 line. 
Starting from the phases (0.60-0.65)\,P, this may indicate the occurrence of gas flows that are directed perpendicularly to the axis of the donor's magnetic field in the direction of the gainer. 
Similar behavior in the phases of (0.1 - 0.6)\,P is demonstrated in the central depth variations of the He\,I $\lambda\lambda$ 3888, 3964 shell lines at the maximum value in the phase of 0.1\,P, and its minimum in the phase of 0.6\,P (see Fig. 14 of \cite{Flora1975}).

It should be noted that all this is consistent with the picture of the directions of the plasma motion of magnetically controlled circumstellar matter in the helium stars according to the study of \cite{Shore1990} (see Figure 11 in the paper of \cite{Shore1990}, and its representation as Fig. 8 in \cite{Skulskyy2020b}). 
Their model can be illustrative for the donor as an oblique magnetic rotator of intermediate obliquity because the dipole axis of its magnetic field deviates relative to the orbital plane of the binary system, and the center of the dipole of the donor's magnetic field is significantly shifted in the direction of the gainer. The outflows of matter from the donor surface in the mass transfer picture are possible both in the direction of the donor's magnetic axis and in the direction perpendicular to this axis \citep[see][]{Skulskyy2020b}. Some proof of this type of phenomenon may be the behavior of a complex line H$_\alpha$  along the surface of the donor. In particular, Fig. 15 of \cite{Flora1975} shows that the equivalent width of the absorption, measured with respect to the total area of the emission in the H$_\alpha$  line, is halved in the phase direction (0.1 - 0.6)\,P, and the total equivalent width of emission plus the absorption of the H$_\alpha$  line shows two deep minima in the phases (0.25-0.45)\,P and about 0.855\,P, i.e. near the phases of visibility of both magnetic poles on the donor surface. Of course, to some extent, this is the result of reflecting the characteristics of the donor that fills its Roche cavity: the loss of matter along the axis of the magnetic field at its pole and the loss of matter at the Lagrange point L1, amplified by the massive gainer and deflected by Coriolis forces to the 0.1\,P phase, are significantly different. At the same time, the variable motions of the plasma  are reflected more clearly in the phases around 0.855\,P of the visibility of the magnetic pole, facing the gainer, in accordance with the changes in the polarity of the magnetic field in the observations of \cite{Skulskij1993}.

\subsection{Magnetic field and the He\,I $\lambda$ 10830 line arising from metastable level}
If the structure and behavior of the He\,I $\lambda$ 3888 line as the second member of the principal series of triplet helium were studied during the first half of the XX century, then to some extent successful observations and studies of the He\,I $\lambda$ 10830 line, as the first member of this series, was carried out by \cite{Alduseva1969} only in 1961-1962. They used a contact image-converter as an image receiver mounted on a 12-inch reflector. The spectral resolution was 4.8 \AA (11 days of observations in 1961) and 2.8 \AA~ at 9 days of observations in 1962. This highlighted the complex structure of the line contour and gave some initial data on the change with the orbital period of the shell radial velocities and equivalent widths of the emission in the He\,I $\lambda$ 10380 line during the orbital period. They indicated that \textquote{the line appears in the very external parts of the shell, surrounding Beta Lyrae as a whole}. \cite{Morgan1974} made the next attempt in the study of the He\,I $\lambda$ 10380 line. They observed on the rapid-scan interferometer at the coudé focus of the 2.7-m telescope at McDonald Observatory in April 1973, but received only two spectra in the 0.90\,P and 0.13\,P phases, respectively. The best resolution of 4.7 \AA~ in the 0.90\,P phase allowed them to record the value of the absorption component -140 km/s with an error of 60 km/s.

\begin{figure}[!t]
	\centerline{\includegraphics[width=0.55\textwidth,clip=]{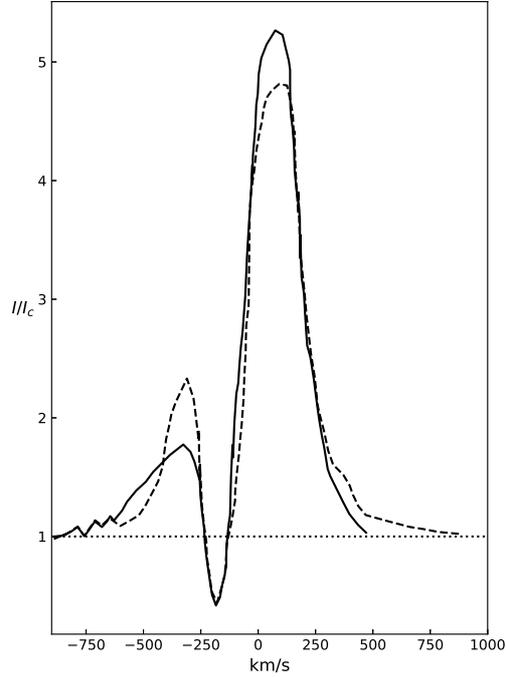}}
	\caption{Contours of the He\,I $\lambda$ 10830 line in Beta Lyrae spectrum in the phases 0.056\,P (solid line) and 0.854\,P (dashed line). Adopted from \cite{Girniak1978}.}
	\label{fig:7}
\end{figure}

Our more complete observations of the He\,I $\lambda$ 10830 line by \cite{Girniak1978} were conducted in April-November 1974. There are no data on the following observations, and these studies need more attention. The observations were performed using an electron-optical image-converter of type FKT-1A (S1) mounted on the 50-inch telescope of the Crimean Astrophysical Observatory. During 26 nights of this season, 59 spectrograms with a dispersion of 48~\AA/mm were obtained, which allowed measuring radial velocities with an error of 10 km/s. As it can be seen from Fig.~\ref{fig:7}, which is a copy of Fig. 1 from \cite{Girniak1978}, this line in the Beta Lyrae spectrum represents the strong emission that is several times higher with respect to the continuum than such emission in the He\,I $\lambda$ 3888 line. The considerable width of the He\,I $\lambda$ 10830 line indicates the movements in the gas plasma structures with velocities up to 800 km/s. The deep absorption component, which cuts through the emission, is shifted to the short-wave side. \cite{Girniak1978} noted that the total intensity and ratio of its short-wavelength and long-wavelength emission components change significantly with the orbital phase. This is a new important factor that requires additional attention and further research.

\begin{figure}[!t]
	\centerline{\includegraphics[width=0.8\textwidth,clip=]{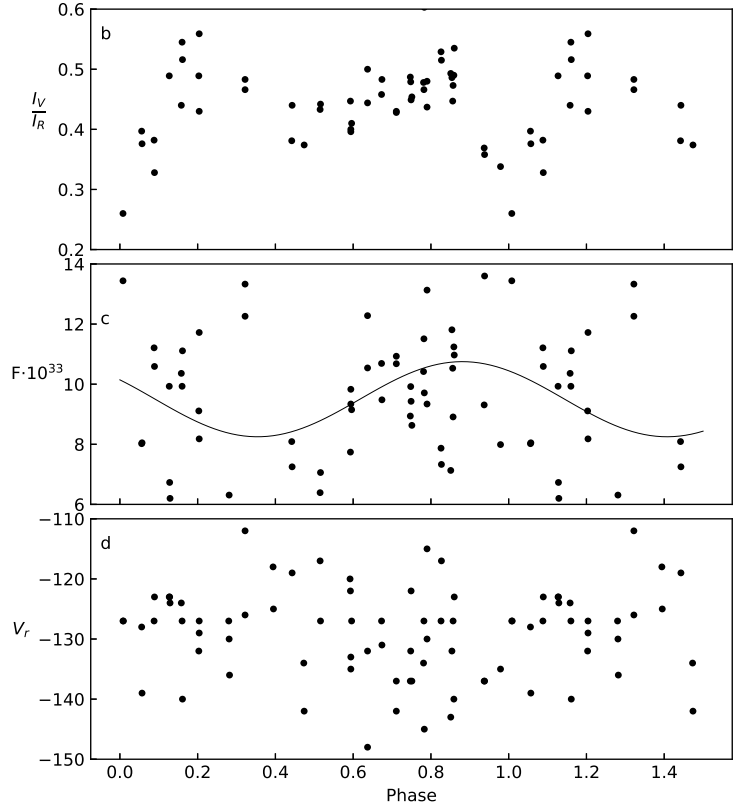}}
	\caption{Changes with the phase of the orbital period of the ratio of intensities of emission components, radiation flux, and radial velocities of the absorption component in the He\,I $\lambda$ 10830 line (plotted according to \cite{Girniak1978}).}
	\label{fig:8}
\end{figure}

\cite{Girniak1978} proposed processed physical parameters of the He\,I $\lambda$ 10830 line, introduced in a general table and five figures. Fig.~\ref{fig:8} reproduces, on the basis of tabular values, two of these five figures,
which are presented below as the radial velocity curve of the absorption component of the He\,I $\lambda$ 10830 line and the sum of its absolute fluxes from violet and red emissions components given in erg\, s$^{-1}$\,A$^{-1}$ km/s, multiplied by $10^{33}$ (contours in the He\,I $\lambda$ 10830 line were built in absolute energy units based on six-color colorimetry \citep{Burnashev1978}). Fig.~\ref{fig:8} indicates some new interesting points. They concern the upper part of Fig. \ref{fig:8}, which illustrates the variability in the orbital phase of the ratio of the intensity of the short-wave and long-wave components of emission, i.e. the dependence of $I_v/I_r = f(P)$, which we built here on the basis of the tabular data. In Fig.~\ref{fig:8}, bottom, it is seen that the $V_r = f(P)$ curve of the absorption component, which separates these emission components, remains approximately the same in the orbital phases at an average value of about -130 km/s (taking into account the radial velocity of -18 km/s for the center of gravity of this binary system). The largest scattering of data and rapid changes in radial velocities are observed in the range of phases (0.7-0.9)\,P, and a particularly sharp velocity change is shown in the phases around 0.855\,P of the visibility of the magnetic pole, facing the gainer. Similar behavior in these phases there is in changes of contours in the He\,I $\lambda$ 10830 line (see the middle part of Fig.\ref{fig:8}). Despite some sharp changes in the total emission, which can occur at times from tens of seconds to tens of minutes (see, e.g., \cite{Alekseev1989} and \cite{Skulskyy2020b}), some trend is observed to a sinusoidal curve of the absolute flux of the total emission with extrema in the phases of (0.35-0.45)\,P and 0.85\,P. It is known that these phases correspond to the phases of both extrema of the donor's magnetic field \citep{Burnashev1991, Skulskij1993}. This picture is more clearly manifested in the dependence of $I_v/I_r = f(P)$ (the upper part of Fig. \ref{fig:8}), characterizing the expected variability of shell structures in this binary system. The dependence of $I_v/I_r = f(P)$ showed a clear decrease in the $I_v/I_r$ ratio after the 0.855\,P phase (of the direct visibility of the magnetic field pole on the donor surface) to the minimum in the main eclipse phase and its rapid rise to the phase 0.15\,P, when the main flow of matter from the donor to the gainer, deflected by the Coriolis forces, begins after the main eclipse. It should be noted that these observations do not provide enough data for more definite conclusions in the phases (0.95-0.05)\,P of the main eclipse.

\begin{figure}[!t]
	\centerline{\includegraphics[width=0.75\textwidth,clip=]{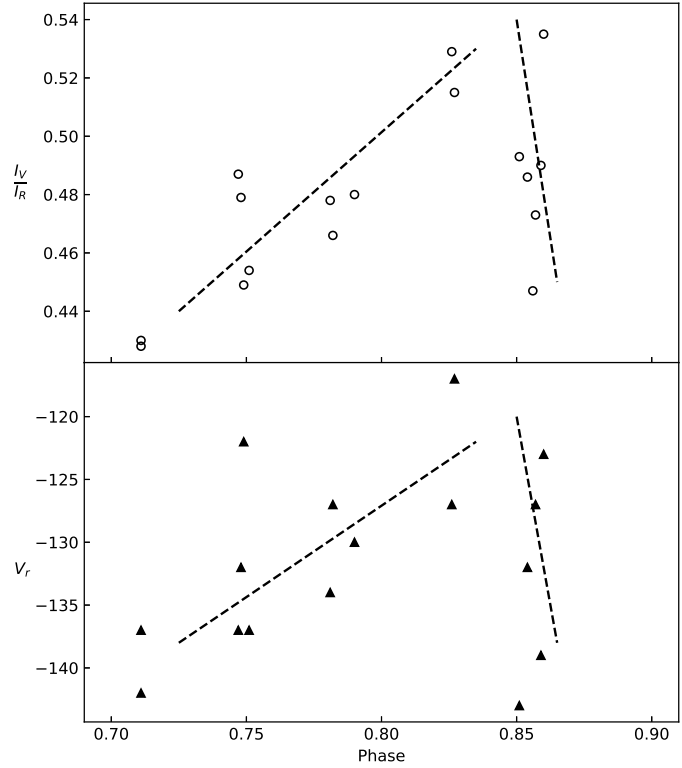}}
	\caption{The same as in Fig. \ref{fig:8} for the He\,I $\lambda$ 10830 line, but on an enlarged scale in phases near 0.855 P of direct visibility of the magnetic pole on the donor surface.}
	\label{fig:9}
\end{figure}

Nevertheless, the radial velocity curve of the He\,I $\lambda$ 10830 line looks similar to that of the He\,I $\lambda$ 3888 line defined by \cite{Skulsky2011} over five different seasons (see Fig.~\ref{fig:5} above). Figs. \ref{fig:5} and \ref{fig:8} show that the average values of the radial velocity of the absorption components of the first two triplet helium lines are close to -130 km/s relative to the center of gravity of the binary system, which significantly exceeds the first parabolic velocity. It could be assumed that the sharp absorption components of both lines are formed in a dynamically identical medium moving in all directions relative to the center of gravity of the binary system. Moreover, the strong emission of the He\,I $\lambda$ 10830 line does not explicitly distort the position of the absorption component, and the radial velocity of the emission center as a whole turned out to be close to the orbital velocity of the gainer, based on the previously established mass ratio of the stellar components \citep{Skulskij1975a}. However, Figs. \ref{fig:1} and \ref{fig:6}, which respectively relate to the studies of \cite{Sahade1959} and \cite{Flora1975}, showed that the behavior of the He\,I $\lambda$ 3888 line at different seasons and epochs can significantly differ. Such facts are real, they required a detailed consideration of some, unnoticed by \cite{Girniak1978}, dynamical characteristics of the He\,I $\lambda$ 10830 line at certain parallels in the study of the lines He\,I $\lambda$ 3888 and He\,I $\lambda$ 10830. 

First of all, this applies to the phase interval (0.75-0.9)\, P around the 0.855\,P phase of the direct visibility of the magnetic field pole on the donor surface close to the gainer. Note that the observations of 1971 by \cite{Flora1975} and our observations of 1974 by \cite{Girniak1978} can be attributed to fairly close seasons and the dynamic behavior of the shell lines He\,I $\lambda$ 3888 and He\,I $\lambda$ 10830 may have some similarity. As it can be seen from the bottom of Fig.~\ref{fig:8}, our observations were most frequent in the (0.7-0.9)\,P phases, showing rapid changes in the radial velocity at large amplitudes of their variability. The observation data at these phases were rebuilt on a larger scale in Fig.~\ref{fig:9}. It can be seen that the radial velocity curve of the He\,I $\lambda$ 10830 line in these phases, like the domed-shaped curve in Fig.~\ref{fig:6} for the He\,I $\lambda$ 3888 line, also has a domed-shaped peak at a clear rise in negative radial velocity immediately after the phase 0.855\,P of direct visibility of the magnetic pole on the donor surface. This peak of the radial velocities of the He\,I $\lambda$ 10830 line that changes in the ranges of 0.1\,P within 20 km/s is twice smaller than that for the He\,I $\lambda$ 3888 in Fig. \ref{fig:6}. At least, this points to some analogies in the dynamic parameters of gas structures in which these triplet helium lines were formed in the first half of the 1970s. The decrease of negative radial velocities at approaching the 0.855\,P phase and their subsequent rapid increase after this phase in the He\,I $\lambda$ 10830 line (as well as in metastable lines of He\,I $\lambda\lambda$  3888, 5015, 3447, and the shell components of the Na\,I $\lambda\lambda$ 5889, 5895 lines - see Fig. \ref{fig:6} built on the data of \cite{Flora1975}) indicates the similarity of physical processes along the donor surface near this magnetic field pole. 

At the same time, the He\,I $\lambda$ 3888 and He\,I $\lambda$ 10830 lines show a physically significant difference in radial velocities. It is clearly seen that if in the He\,I $\lambda$ 10830 line in Fig. \ref{fig:9} in all phases, including phases (0.6-0.8)\,P, the radial velocities are greater than the parabolic velocity, then in the He\,I $\lambda$ 3888 line according to Fig.\ref{fig:6} in phases (0.6-0.8)\,P, they are much smaller than the parabolic velocity. This indicates the nonstationarity of the outflow of matter and the variable flow density in the direction of the dipole axis of the magnetic field. The clear changes, which were noticed in the absolute radiation flux (the central part of Fig. \ref{fig:8}) within (0.8-0.9)\,P, also reflect the complex structure of the emission formation near 0.855\,P in this short phase interval with a width of approximately 0.1\,P. This emphasizes that rapid changes are inherent in these phases of both energy and dynamic parameters of the magnetized plasma. The rapid change in radial velocities near 0.855\,P may indicate that the vector components of the magnetized plasma motion at the edges of the magnetic field pole are superimposed on the total plasma motion in the direction of the donor magnetic field axis. This one can also see from Fig.\ref{fig:4}, which demonstrates similar variations in the radial velocities of the He\,I $\lambda\lambda$ 3964, 5015 lines in the short phase interval with a width of approximately 0.1\,P in phases of the visibility of both magnetic field poles on the donor surface. The observations of \cite{Girniak1978} in phases (0.25-0.45)\,P are not dense enough to conduct a detailed analysis of the behavior of the He\,I $\lambda$ 10830 line in the phases of visibility of the magnetic field pole on the donor surface in the first quadrature, although, in these phases, the radial velocities of this line are also subject to abrupt changes similar to Fig.\ref{fig:4}. However, as a whole, the above research facts confirm the general picture of the behavior of energetics and dynamic parameters of magnetized plasma caused by a certain configuration of the donor's magnetic field. The study of the spectral characteristics of the He\,I $\lambda$ 10830 line has led to certain clarifications in the formation of the picture of the transfer matter between stellar components and its movement into the medium around the Beta Lyrae system.

\section{Conclusions and discussion}
The study of shell structures in the Beta Lyrae system seemed natural on the basis of previous studies by \cite{Skulskyy2020a, Skulskyy2020b}, which involved a continuous interaction of the outer shell with a complex system of plasma flows between the donor and the gainer. It is clear that the spectral lines formed in external structures, in particular the helium lines arising from metastable levels, were studied. Scientific articles were used for the study, which, on the basis of qualitative observations, covered the orbital period well. There are few similar works, in particular \cite{Sahade1959, Flora1975, Girniak1978}, which are half a century old. It is to some extent well that the authors of these works presented their results without the influence of modern interpretations, say, without information about the magnetic field of the donor or ideas about bipolar jet structures associated with the accretion disk. However, these articles have a lot of important spectral material in the form of published tables and graphs. As a whole, this material, together with data from more recent publications of \cite{Harmanec1993} and \cite{Skulsky2011}, have not been analyzed in more detail. Here an attempt was made to overcome this shortcoming.

All researched articles are distinctive. They differ significantly in the energy and dynamic parameters of the moving plasma in the medium near each component and between stellar components of the binary system. First, significant long-term differences were found within half a century in the movements of matter to the gainer and in external structures. Second, there are significant differences in the physical parameters of this plasma within one observation season and their variations from cycle to cycle (especially in \cite{Sahade1959}). It was established (see Fig.~\ref{fig:1}) only in \cite{Sahade1959} that: \textquote{He\,I $\lambda$ 3888 has two components with very definite and fairly constant velocities} of about -155 km/s and -115 km/s (in the 1955 season), i.e., close to the second and first parabolic velocities. In \cite{Flora1975}, the absorption component of the He\,I $\lambda$ 3888 line in certain phases of the 1971 season is approximately -80 km/s, which is much less than the parabolic velocity (see Fig. \ref{fig:6}). The averaged radial velocity of the absorption component in the He\,I $\lambda$ 3888 line during the 1991 season is -127 km/s (see also Fig.3 in \cite{Harmanec1993}), which is very different from that in the previous two articles. The He\,I $\lambda \lambda$  3964, 5015 lines (Fig. \ref{fig:4}) show rapid changes in radial velocities in the range from -55 km/s to -125 km/s in short phase intervals up to 0.1\,P, in particular close to the magnetic field poles, which were recorded in the seasons 2008-2010 in \cite{Skulsky2011}. It seems that all these differences are not periodically ordered. It should be emphasized that all researched articles show the averaged radial velocity in the He\,I $\lambda \lambda$  3888, 10830, i.e., the velocity of the outflowed plasma in external structures, to be of about -130 km/s, except for one orbital cycle in 1971, in which observations by \cite{Flora1975} were made. 

The article gradually reveals the analyzed facts from the available spectral material, and since there is quite a substantial amount of them, they are further presented as touches of the obtained research picture in a possibly concise form. It turned out that this picture is based on fragments of a generalized investigation during the orbital period. They can be grouped into certain phase intervals as a reflection of slightly different physical processes occurring in them. Formally, they can be divided into five interconnected intervals. The phase range (0.95-0.05)\,P reflects nonstationary processes that occur during the transfer of mass to the gainer at the donor eclipse by the accretion disk. The phase range (0.07-0.14)\,P reflects the motion of the main gas flow deflected by Coriolis forces outside the accretion disk. It is necessary to distinguish the behavior of plasma structures near the phases of visibility of both poles of the magnetic field (0.30-0.45)\,P and (0.75-0.90)\,P, as well as within the range of phases (0.4-0.6)\,P eclipse of the accretion disk by the donor. 

There are a number of effects and events associated with the eclipsing of the donor surface by the accretion disk in the phases (0.9-0.1)\,P of the main eclipse of the binary system. It can be noted that within phases (0.95-0.05)\,P of the deeper eclipse, the He\,I $\lambda \lambda$ 3614, 3964 and He\,I $\lambda$ 3888 absorption lines show, according to \cite{Sahade1959}, two main regions of changes in their intensities and radial velocities. They are grouped near the 0.95\,P and 0.05\,P phases, which corresponds to the phases of maximum accretion disk density. This follows from many studies, in particular from phase changes of the equivalent widths of satellite lines in 1991-1992 (see Fig. 2 in \cite{Skulskij1993c}, or from the light curve based on observations by the BRITE Satellites in May-October 2016 (see Fig. 2 in \cite{Rucinski2018}). In these phases, there is the greatest variability of the radial velocities of both absorption components of the He\,I $\lambda \lambda$ 3614, 3965 and He\,I $\lambda$ 3888 lines. Their more negative shifted component varies in the range from -140 km/s to -160 km/s, sometimes reaching the second parabolic velocity (near -160 km/s) without a certain regularity in orbital cycles. This also applies to the center of the eclipse in different orbital cycles in phases (0.98-0.02)\,P, where the radial velocities also show variation within the same limits (for example, in phases (0.98-0.99)\,P at the end of the 6th cycle on July 4, 1955, their radial velocity was -(155-160) km/s, that is sufficient for an outflow matter from the Lagrange point L3). It is clear that in the range (0.95-0.05) \,P the dynamic behavior of moving matter reflected in the more negative absorption component in the He\,I $\lambda \lambda$ 3614, 3964 and He\,I $\lambda$ 3888 lines can be the product of instability processes of the matter transfer between the stellar components. The less negative absorption component of the He\,I $\lambda$ 3888 line in these phases reaches only the first parabolic velocity, and in the He\,I $\lambda \lambda$ 3614, 3964 lines they only approach it or are much smaller, also demonstrating the fact of non-laminar, but clearly variable events during the mass transfer. It should be noted that the observations of \cite{Sahade1959} were carried out in the 1955 season with significant phase gaps during the day (its time is 0.077\,P), which does not allow to clarify the structure of the accretion disk. The important fact of the detection of disk-shaped structures in phases (0.95-0.05)\,P are the observations of \cite{Skulsky2011} during 2008-2010, which clearly show rapid changes of radial velocities from -55 km/s to -125 km/s in He\,I $\lambda \lambda$ 3965, 5015 singlet lines (see Fig. \ref{fig:4}) in two short phase intervals.

Thus, the variable nature of the flows of matter during mass transfer in phases of the main eclipse indicates eruption processes on the donor, the appearance of matter clots over the rims of the disk from the side of the donor and their movement in the direction to the observer over the surface of the accretion disk accelerated by the massive gainer. Incidentally, \cite{Skulskyy2020b} showed that the vertical component of the plasma flows in the direction of the gainer can be formed due to the configuration of the donor's magnetic field, in which the magnetic axis is inclined at a significant angle to the orbit of the binary system. The structure of gas flows above the surface of the accretion disk significantly depends on the location of the magnetic pole on the donor surface relative to its distance to the gainer and relative to the plane of the orbit. These remarks also apply to the third region of active events in the phases (0.9-0.1)\,P reflected in the changeable and denser flows of matter in the phases (0.07-014)\,P, which start after the center of the main eclipse of the binary system. The dynamic behavior of the moving matter in the He\,I $\lambda \lambda$ 3614, 3964 and He\,I $\lambda$ 3888 lines is similar to that in the phase range (0.95-0.05)\,P. Interestingly, the highest radial velocity for the observation season in \cite{Sahade1959} was recorded in phases (0.11-0.13)\,P of the 5th cycle on June 10, 1955. The more negative absorption component of the He\,I $\lambda$ 3888 line showed the velocity of -(160-170) km/s, which is more than enough for the outflow of the matter from the Lagrange point L3. These more dense gas flows, when deflected by Coriolis forces outside the accretion disk in the Roche cavity of the gainer, push the outer part of the disk to the gainer surface \citep[see][]{Skulskij1992, Skulskyy2020a}.

The range of phases (0.4-0.6)\,P eclipse of the accretion disk by the donor, as a specific part of the orbital period, deserves a separate study (by analogy with the studies in phases (0.9-0.1)\,P of the primary eclipse). The preliminary study of existing observations in this phase range leads to certain conclusions and ideas about the behavior of physical parameters in the vicinity of both stellar components. Although they require additional observations, they were worth paying attention to. Analysis of Tables 3 and 4 together with their illustrations 13-16 in the article \cite{Sahade1959} (see also Fig. \ref{fig:1}-\ref{fig:3} above), as well as the study of \cite{Skulskij1992, Skulskij1993a, Skulskij1993b, Skulskij1993c, Ak2007} and others, leads to the conclusion that in these phases, in addition to complex shell lines formed near the accretion disk, clear changes of the curve of radial velocities of the gainer lines are visible. Their behavior can be explained by the Rossiter effect in the passage of the donor in front of the accretion disk. In phases (0.38-0.50)\,P the right part of the disk is observed (its left part being closed by the donor); in phases (0.45-0.55)\,P the central parts of the disk, closed by the donor, pass in front of the observer; in phases (0.5-0.6)\,P dominates in these lines the left part of the accretion disk (the donor obscures the right part of the disk). Approximately at 0.62\,P, this effect disappears (further there are opened phases of the collision of the main flow into the accretion disk and the region of the magnetic field pole on the donor surface). In particular, this effect is well observed in phases (0.38-0.62)\,P in the behavior of the gainer radial velocities of the Si\,II $\lambda \lambda$ 6347, 6371 lines, for example, in Fig. 3 in \cite{Skulskij1992}, where it was revealed, and, especially in Fig. 5 in \cite{Ak2007}. It is more likely that these lines are formed near the central part of the disk.

The visibility phases of both poles of the magnetic field (0.30-0.45)\,P and (0.75-0.90)\,P as specific parts of the orbital period are important for obtaining a general picture of the influence of the donor's magnetic field on moving magnetized gas structures in near and farther medium regarding the donor. Some of the key facts identified in this study are, first of all, illustrated by the behavior of the radial velocities of helium lines arising from metastable levels. Fig. \ref{fig:1} shows that only within phases of about 0.1\,P near the phases 0.855\,P, both absorption components of the He\,I $\lambda$ 3888 line in \cite{Sahade1959} converge into a single absorption with radial velocities in the range -(125-135) km/s, i.e., with a radial velocity much greater than the first parabolic velocity. This indicates that in the phase of direct visibility of the pole of the magnetic field on the surface of the donor facing the gainer, the matter outflows in the direction of the axis of the donor magnetic field, perpendicular to the surface of the donor. According to Fig.~\ref{fig:3}, the radial velocities from the shells of the He\,I $\lambda \lambda$ 3614, 3964 singlet lines, although showing some stratification of their formation relative to the line He\,I $\lambda$ 3888, indicate that the He\,I lines of both series in phases close to 0.855\,P behave in these phases in a similar way. The particularly sharp peaks of radial velocities with a rapid change in their values are observed in Fig.~\ref{fig:4} in the He\,I $\lambda \lambda$ 3964, 5015 lines according to \cite{Skulsky2011}. Within 0.1\,P in the 0.355\,P and 0.855\,P phases, there are sharp changes in the radial velocity from -55 km/s to -125 km/s, or 70 km/s relative to the center of gravity of the binary system. This indicates complex processes in the plasma structures of this binary system, which are close to the location of the magnetic poles on the donor surface. It should be recognized that Fig.~\ref{fig:9} also confirms that the rapid changes in the radial velocity of the He\,I $\lambda$ 10830 line in phases immediately close to 0.855\,P (at the mean velocities over the orbital period near -130 km/s - see Fig.~\ref{fig:8}) indicates the reality of rapid changes in plasma movement and events in this phase range. 

Evidencing certain stratification of their formation, the He\,I lines of both series show in theses phases a certain relationship between their dynamic and energy parameters. Figure~\ref{fig:9} also shows the parallel changes in the profile of the He\,I $\lambda$ 10830 line expressed through the dependence of $I_v/I_r = f(P)$, i.e., the ratio of the intensity of the short-wave and long-wave components of these emission profiles. This dependence is similar to the radiation increasing near the 0.855\,P phase on light curves in the range of (3.5-4.6) m in \cite{Zeilik1982}, which allowed to identify \citep[see][]{Skulskyy2020b} on the donor surface the region of this magnetic field pole. Figure~\ref{fig:2} illustrates also the radial velocity curve from the red emission peak at the He\,I $\lambda$ 3888 line. Such two maxima are observed on the radial velocity curve of the red peak at the H$_\alpha$ line (see Figures~5 in \cite{Skulskyy2020b}. The maxima at the 0.355\,P and 0.855\,P phases coincide with the phases of the two maxima on the curve of the effective magnetic field strength of the donor (see Figure~1 in \cite{Skulskyy2020b} or Figure~2 in \cite{Burnashev1991}). Moreover, the radial velocity curve for the Gaussian center of emission in H$_\alpha$, as well as for such centers on radial velocity curves of the He\,I $\lambda \lambda$ 6678, 7065, and Si\,II $\lambda \lambda$ 6347, 6371 lines, see Figure~6 in \cite{Skulskyy2020b}, demonstrate that both their maxima match the extrema of the effective magnetic field strength of the donor (a similar shift on the $V_r$-curve of the center of the He\,I $\lambda$ 10830 emission as a whole is discussed by \cite{Girniak1978}). The generation of the radiation flux and the formation of the emission-absorption profile of these lines can be largely carried out under the influence of the existing configuration of the donor magnetosphere in the interconnected spatial structures.

A similar conclusion is supported by the observations of \cite{Flora1975}. Conducted within one orbital cycle, they showed unusual variability of physical parameters within the phases (0.70-0.90)\,P close to the 0.855\,P phase of the magnetic field pole close to the gainer. Fig.~\ref{fig:6} shows a much lower velocity of the moving plasma than the first parabolic velocity in all five helium lines arising from metastable levels. This clearly indicates  nonstationarity of the outflow of matter and the variable flow density in the direction of the dipole axis of the magnetic field. The behavior of radial velocities in phases (0.8-0.9)\,P in the shell lines Na\,I $\lambda \lambda$  5889, 5895 is also particularly important. During 0.1\,P, the moving plasma slows down, stops at a certain point, and accelerates again, stating the fact of actual variations in the sign of this movement. Thus, the study of observations of \cite{Flora1975} also confirmed that the observer, looking perpendicular to the surface of the donor along the axis of the magnetic field, reveals the region of the donor surface that is responsible for the location of the pole of the magnetic field facing the gainer and the loss of matter directly from the donor surface in phase 0.855\,P with its subsequent transfer to the gainer.

However, the fact of localization of the poles of the magnetic field on the surface of the donor, reflected in many studies, needs some clarification, based on the spatial location of its magnetic axis. Let us turn to the model of the donor as a magnetic rotator, by analogy to the model of a helium star as an oblique magnetic rotator presented by \cite{Shore1990}. This model (see Fig. 8 in \cite{Skulskyy2020b}) may be quite illustrative since the dipole axis of its magnetic field is deviated by 28\degr~relative to the orbital plane of the binary system \citep{Skulskij1985}. The center of the donor magnetic dipole is displaced by 0.08 of the distance between the centers of gravity of both components toward the gainer's center. It is also clear that the magnetic pole, located on the surface of the donor and observed in phases of about 0.855\,P, is more effective in terms of the amount of transferred matter to the gainer, which is confirmed by all studies. The ionized gas, directed by the magnetic field of the donor in the direction of its dipole axis from the surface of the donor, is deflected along the lines of the magnetic field primarily to the accretion disk. However, a certain amount of charged particles will move along the lines of force of the magnetic field in the direction to the second pole of the magnetic field on the surface of the donor, heating its surface. And, if in the first quadrature in the phases around 0.355\,P the magnetic pole on the surface of the donor is located above the plane of the orbit, then in the phases around 0.855\,P the magnetic pole on the surface of the donor is below the plane of the orbit, or vice versa. The detection of these poles also depends on the inclination of the plane of the orbit. Not everything is clear in this part because the inclination of the orbit in current research is taken from i = 81\degr~ in \cite{Mennickent2013} to i = 93.5\degr~ in \cite{Mourard2018}. This is close to the binary orbital inclination i = 90\degr, and regardless of whether the pole of the magnetic field is in the upper or lower hemisphere on the donor surface, the observer, due to the projection of the rounded shape of the surface on the line of sight, must register certain deviations of the surface heating maximum from the 0.355\,P phase. This is confirmed by the absolute spectrophotometry of \cite{Burnashev1991} of 1974-1985. They showed that the observer starts to register the excess radiation (a hot spot on the donor surface) only from the 0.355\,P phase directed along the axis of the magnetic field (see Figure~1 in \cite{Skulskyy2020b}). The excess of this radiation disappears near the 0.50\,P phase, demonstrating during the orbital phases (0.37-0.49)\,P the rapid variability of the absolute radiation flux in the H$_\alpha$ emission line and the continuum around this line \citep[see][]{Skulskyy2020b, Alekseev1989}. The maximum of the excess of this radiation is really shifted from the 0.355\,P phase and corresponds to the phases (0.43-0.47)\,P. Thus, in these phases, some surface heating on the donor surface may be formed due to nonstationary shock collisions of ionized gas, directed along the magnetic field lines to the magnetic field pole, the location of which is given by the spatial configuration of the magnetic field dipole. The phase range (0.43-0.47)\,P corresponds to the known minimum on the polarization curves for Beta Lyrae studied by \cite{Appenzeller1967} and \cite{Lomax2012}, which is interpreted as formed by collisions of gas flows with the accretion disk during the scattering of radiation by free electrons. This encourages the research of the reflection of the spatial configuration of the donor's magnetic field in the polarization observations.

\acknowledgements
The author is thankful to V.I. Kudak for consultations.

\bibliography{Skulsky}

\begin{thebibliography}{31}
\expandafter\ifx\csname natexlab\endcsname\relax\def\natexlab#1{#1}\fi

\bibitem[{{Ak} {et~al.}(2007){Ak}, {Chadima}, {Harmanec}, {Demircan}, {Yang},
  {Koubsk{\'y}}, {{\v{S}}koda}, {{\v{S}}lechta}, {Wolf}, {Bo{\v{z}}i{\'c}},
  {Ru{\v{z}}djak}, \& {Sudar}}]{Ak2007}
{Ak}, H., {Chadima}, P., {Harmanec}, P., {et~al.}, {New findings supporting the
  presence of a thick disc and bipolar jets in the {\ensuremath{\beta}} Lyrae
  system}. 2007, {\it \aap}, {\bf 463}, 233, DOI: 10.1051/0004-6361:20065536

\bibitem[{{Alduseva} \& {Esipov}(1969)}]{Alduseva1969}
{Alduseva}, V.~Y. \& {Esipov}, V.~F., {The line He\,I {\ensuremath{\lambda}}
  10830 in {\ensuremath{\beta}} Lyr shell}. 1969, {\it \azh}, {\bf 46}, 113

\bibitem[{{Alexeev} \& {Skulskij}(1989)}]{Alekseev1989}
{Alexeev}, G.~N. \& {Skulskij}, M.~Y., {Rapid variability of the spectrum of
  {\ensuremath{\beta}} Lyrae in the {\ensuremath{H_\alpha}} region}. 1989, {\it
  Bull. Spec. Astroph. Obs.}, {\bf 28}, 21

\bibitem[{{Appenzeller} \& {Hiltner}(1967)}]{Appenzeller1967}
{Appenzeller}, I. \& {Hiltner}, W.~A., {True polarization curves for Beta
  Lyrae}. 1967, {\it \apj}, {\bf 149}, 353, DOI: 10.1086/149258

\bibitem[{{Bisikalo} {et~al.}(2000){Bisikalo}, {Harmanec}, {Boyarchuk},
  {Kuznetsov}, \& {Hadrava}}]{Bisikalo2000}
{Bisikalo}, D.~V., {Harmanec}, P., {Boyarchuk}, A.~A., {Kuznetsov}, O.~A., \&
  {Hadrava}, P., {Circumstellar structures in the eclipsing binary
  {\ensuremath{\beta}} Lyr A. Gasdynamical modelling confronted with
  observations}. 2000, {\it \aap}, {\bf 353}, 1009

\bibitem[{{Burnashev} \& {Skulskij}(1978)}]{Burnashev1978}
{Burnashev}, V.~I. \& {Skulskij}, M.~Y., {Absolute spectrophotometry of
  {\ensuremath{\beta}} Lyr.} 1978, {\it \krym}, {\bf 58}, 64

\bibitem[{{Burnashev} \& {Skulskij}(1991)}]{Burnashev1991}
{Burnashev}, V.~I. \& {Skulskij}, M.~Y., {H$_{{\ensuremath{\alpha}}}$
  photometry and magnetic field of {\ensuremath{\beta}} lyrae}. 1991, {\it
  \krym}, {\bf 83}, 108

\bibitem[{{Flora} \& {Hack}(1975)}]{Flora1975}
{Flora}, U. \& {Hack}, M., {Spectrographic observations of {\ensuremath{\beta}}
  Lyr during the international campaign of 1971.} 1975, {\it \aaps}, {\bf 19},
  57

\bibitem[{{Girnyak} {et~al.}(1978){Girnyak}, {Skulskij}, {Shanin}, \&
  {Shcherbakov}}]{Girniak1978}
{Girnyak}, M.~B., {Skulskij}, M.~Y., {Shanin}, G.~I., \& {Shcherbakov}, A.~G.,
  {The investigation of the emission line of He\,I {\ensuremath{\lambda}} 10830
  {\r{A}} in the spectrum of Beta Lyrae.} 1978, {\it \krym}, {\bf 58}, 75

\bibitem[{{Harmanec} \& {Scholz}(1993)}]{Harmanec1993}
{Harmanec}, P. \& {Scholz}, G., {Orbital elements of {\ensuremath{\beta}} Lyrae
  after the first 100 years of investigation.} 1993, {\it \aap}, {\bf 279}, 131

\bibitem[{{Hoffman} {et~al.}(1998){Hoffman}, {Nordsieck}, \&
  {Fox}}]{Hoffman1998}
{Hoffman}, J.~L., {Nordsieck}, K.~H., \& {Fox}, G.~K., {Spectropolarimetric
  evidence for a bipolar flow in Beta Lyrae}. 1998, {\it \aj}, {\bf 115}, 1576,
  DOI: 10.1086/300274

\bibitem[{{Lomax} {et~al.}(2012){Lomax}, {Hoffman}, {Elias}, {Bastien}, \&
  {Holenstein}}]{Lomax2012}
{Lomax}, J.~R., {Hoffman}, J.~L., {Elias}, Nicholas~M., I., {Bastien}, F.~A.,
  \& {Holenstein}, B.~D., {Geometrical constraints on the hot spot in Beta
  Lyrae}. 2012, {\it \apj}, {\bf 750}, 59, DOI: 10.1088/0004-637X/750/1/59

\bibitem[{{Mennickent} \& {Djura{\v{s}}evi{\'c}}(2013)}]{Mennickent2013}
{Mennickent}, R.~E. \& {Djura{\v{s}}evi{\'c}}, G., {On the accretion disc and
  evolutionary stage of {\ensuremath{\beta}} Lyrae}. 2013, {\it \mnras}, {\bf
  432}, 799, DOI: 10.1093/mnras/stt515

\bibitem[{{Morgan} {et~al.}(1974){Morgan}, {Potter}, \& {Kondo}}]{Morgan1974}
{Morgan}, T.~H., {Potter}, A.~E., \& {Kondo}, Y., {Complex infrared emission
  features in the spectrum of Beta Lyrae.} 1974, {\it \apj}, {\bf 190}, 349,
  DOI: 10.1086/152883

\bibitem[{{Mourard} {et~al.}(2018){Mourard}, {Bro{\v{z}}}, {Nemravov{\'a}},
  {Harmanec}, {Budaj}, {Baron}, {Monnier}, {Schaefer}, {Schmitt},
  {Tallon-Bosc}, {Armstrong}, {Baines}, {Bonneau}, {Bo{\v{z}}i{\'c}},
  {Clausse}, {Farrington}, {Gies}, {Jury{\v{s}}ek}, {Kor{\v{c}}{\'a}kov{\'a}},
  {McAlister}, {Meilland}, {Nardetto}, {Svoboda}, {{\v{S}}lechta}, {Wolf}, \&
  {Zasche}}]{Mourard2018}
{Mourard}, D., {Bro{\v{z}}}, M., {Nemravov{\'a}}, J.~A., {et~al.}, {Physical
  properties of {\ensuremath{\beta}} Lyrae A and its opaque accretion disk}.
  2018, {\it \aap}, {\bf 618}, A112, DOI: 10.1051/0004-6361/201832952

\bibitem[{{Rucinski} {et~al.}(2018){Rucinski}, {Pigulski}, {Popowicz},
  {Kuschnig}, {Koz{\l}owski}, {Moffat}, {Pavlovski}, {Hand ler}, {Pablo},
  {Wade}, {Weiss}, \& {Zwintz}}]{Rucinski2018}
{Rucinski}, S.~M., {Pigulski}, A., {Popowicz}, A., {et~al.}, {Light-curve
  instabilities of {\ensuremath{\beta}} Lyrae observed by the BRITE
  satellites}. 2018, {\it \aj}, {\bf 156}, 12, DOI: 10.3847/1538-3881/aac38b

\bibitem[{{Sahade} {et~al.}(1959){Sahade}, {Huang}, {Struve}, \&
  {Zebergs}}]{Sahade1959}
{Sahade}, J., {Huang}, S.~S., {Struve}, O., \& {Zebergs}, V., The spectrum of
  Beta Lyrae. 1959, {\it Transactions of the American Philosophical Society},
  {\bf 49}, 1

\bibitem[{{Shore} \& {Brown}(1990)}]{Shore1990}
{Shore}, S.~N. \& {Brown}, D.~N., {Magnetically controlled circumstellar matter
  in the helium-strong stars}. 1990, {\it \apj}, {\bf 365}, 665, DOI:
  10.1086/169520

\bibitem[{{Skulskij}(1975)}]{Skulskij1975a}
{Skulskij}, M.~Y., {Quantitative analysis of the spectrum of Beta Lyrae IV.
  Line identifications for the faint component and the mass of both stars}.
  1975, {\it \azh}, {\bf 52}, 710

\bibitem[{{Skulskij}(1985)}]{Skulskij1985}
{Skulskij}, M.~Y., {The magnetic field of the Beta-Lyrae system}. 1985, {\it
  \sal}, {\bf 11}, 21

\bibitem[{Skulskij(1992)}]{Skulskij1992}
Skulskij, M.~Y., Study of {\ensuremath{\beta}} Lyrae CCD spectra. Absorbtion
  lines, orbital elements and disk structure of the gainer. 1992, {\it \sal},
  {\bf 18}, 287

\bibitem[{{Skulskij}(1993{\natexlab{a}})}]{Skulskij1993b}
{Skulskij}, M.~Y., {Spectra of {\ensuremath{\beta}} Lyr. Matter transfer and
  circumstellar structures in presence of the donor’s magnetic field}.
  1993{\natexlab{a}}, {\it Astron. Lett.}, {\bf 19}, 45

\bibitem[{{Skulskij}(1993{\natexlab{b}})}]{Skulskij1993a}
{Skulskij}, M.~Y., {Study of {\ensuremath{\beta}} Lyrae spectra - the Si\,II
  {\ensuremath{\lambda \lambda}}\,6347, 6371 doublet and the discovery of the
  cyclic variability of equivalent widths of lines of the loser's "magnetized"
  atmosphere}. 1993{\natexlab{b}}, {\it Astron. Lett.}, {\bf 19}, 19

\bibitem[{{Skulskij}(1993{\natexlab{c}})}]{Skulskij1993c}
{Skulskij}, M.~Y., {The spectrum of {\ensuremath{\beta}} Lyrae: the SiII
  {\ensuremath{\lambda \lambda}}\,6347, 6371 doublet in 1992 and its variation
  from season to season}. 1993{\natexlab{c}}, {\it Astron. Lett.}, {\bf 19},
  160

\bibitem[{{Skulskij} \& {Plachinda}(1993)}]{Skulskij1993}
{Skulskij}, M.~Y. \& {Plachinda}, S.~I., {A study of the magnetic field of the
  bright component of {\ensuremath{\beta}} Lyr in the SiII {\ensuremath{\lambda
  \lambda}}\,6347, 6371 lines}. 1993, {\it Pisma Astron. Zh.}, {\bf 19}, 517

\bibitem[{{Skulsky} \& {Kos}(2011)}]{Skulsky2011}
{Skulsky}, M.~Y. \& {Kos}, E.~S., {On the dynamics of circumstellar gaseous
  structures and magnetic field of {\ensuremath{\beta}} Lyrae}. 2011, in {\it
  Magnetic Stars. Proceedings of the International Conference, held in the
  Special Astrophysical Observatory of the Russian AS, August 27- September 1,
  2010}, ed. I.~{Romanyuk} \& D.~{Kudryavtsev}, 259--263

\bibitem[{{Skulskyy}(2020{\natexlab{a}})}]{Skulskyy2020a}
{Skulskyy}, M.~Y., {Formation of magnetized spatial structures in the Beta
  Lyrae system. I. Observation as a research background of this phenomenon}.
  2020{\natexlab{a}}, {\it \caos}, {\bf 50}, 681, DOI:
  10.31577/caosp.2020.50.3.681

\bibitem[{{Skulskyy}(2020{\natexlab{b}})}]{Skulskyy2020b}
{Skulskyy}, M.~Y., {Formation of magnetized spatial structures in the Beta
  Lyrae system. II. Observation as a research background of this phenomenon}.
  2020{\natexlab{b}}, {\it \caos}, {\bf 50}, 717, DOI:
  10.31577/caosp.2020.50.4.717

\bibitem[{{Struve}(1941)}]{Struve1941}
{Struve}, O., {The Spectrum of {\ensuremath{\beta}} Lyrae.} 1941, {\it \apj},
  {\bf 93}, 104, DOI: 10.1086/144249

\bibitem[{{Umana} {et~al.}(2000){Umana}, {Maxted}, {Trigilio}, {Fender},
  {Leone}, \& {Yerli}}]{Umana2000}
{Umana}, G., {Maxted}, P.~F.~L., {Trigilio}, C., {et~al.}, {Resolving the radio
  nebula around Beta Lyrae}. 2000, {\it \aap}, {\bf 358}, 229

\bibitem[{{Zeilik} {et~al.}(1982){Zeilik}, {Heckert}, {Henson}, \&
  {Smith}}]{Zeilik1982}
{Zeilik}, M., {Heckert}, P., {Henson}, G., \& {Smith}, P., {Infrared photometry
  of Beta Lyrae: 1977-1982.} 1982, {\it \aj}, {\bf 87}, 1304, DOI:
  10.1086/113217

\end{thebibliography}

\end{document}